\documentclass[twocolumn]{aastex63}
\usepackage{amssymb,amsmath}
\usepackage{graphicx}
\usepackage{xcolor,natbib}
\usepackage{multirow}
\usepackage[T1]{fontenc}
\usepackage{ae,aecompl}
\usepackage{newtxtext, newtxmath}
\usepackage{float}
\usepackage{subfigure}
\usepackage{longtable}
\usepackage{enumitem}
\usepackage{longtable,listings}
\usepackage[flushleft]{threeparttable}
\usepackage{parcolumns}
\usepackage{textcomp}

\def\oii{[O~{\sc ii}]$\lambda3727$\AA\ }

\def\o3{[O~{\sc iii}]}

\submitjournal{ApJS}

\shorttitle{NLRs sizes of BLAGN}
\shortauthors{ZHANG}

\begin{document}

\title{Dependence of NLRs sizes on \o3~ luminosity in low redshift AGN with double-peaked 
broad Balmer emission lines}

\correspondingauthor{XueGuang Zhang}
\email{xgzhang@njnu.edu.cn}

\author{Zhang XueGuang}
\affiliation{School of physics and technology, Nanjing Normal University,
                No. 1, Wenyuan Road, Nanjing, Jiangsu, 210046, P. R. China}

\begin{abstract} 
In the manuscript, the simple but interesting results are reported on the upper limits of NLRs 
sizes of a small sample of 38 low redshift ($z<0.1$) AGN with double-peaked broad 
emission lines (double-peaked BLAGN), in order to check whether the NLRs sizes in type-1 AGN 
and type-2 AGN obey the similar empirical dependence on \o3~ luminosity. In order to correct 
the inclination effects on projected NLRs sizes of type-1 AGN, the accretion disk origin is 
commonly applied to describe the double-peaked broad H$\alpha$ leading to the determined 
inclination angles of central disk-like BLRs of the 38 double-peaked BLAGN. Then, 
considering the fixed SDSS fiber radius, the upper limits of NLRs sizes of the 38 
double-peaked BLAGN can be estimated. Meanwhile, the strong linear correlation between continuum 
luminosity and \o3~ luminosity is applied to confirm that the \o3~ emissions of the 38 
double-peaked BLAGN are totally covered in the SDSS fibers. Considering the reddening corrected 
measured \o3~ luminosity, the upper limits of NLRs sizes of the 38 double-peaked 
BLAGN are within the 99.9999\% confidence interval of the expected results from the empirical 
relation between NLRs sizes and \o3~ luminosity in the type-2 AGN. In the current stage, there 
are no clues to challenge the Unified model of AGN, through the space properties of NLRs.
\end{abstract}


\section{Introduction}

	Strong \o3~ emission lines are fundamental characteristics of both broad line Active Galactic 
Nuclei (type-1 AGN) and narrow line AGN (type-2 AGN). Under the current framework of the preferred 
being-improved Unified Model \citep{an93, nh15}, type-1 AGN and type-2 AGN have totally the same 
central structures of narrow emission line regions (NLRs) and broad emission line regions (BLRs) (the 
two essential structures of AGN), but the BLRs are seriously obscured by central dust torus leading 
to no apparent broad optical emission lines detected in type-2 AGN. The well detected polarized broad 
emission lines in type-2 AGN, such as the results well discussed in \citet{am85, tr01, tr03, nk04, 
cm07, sg10, bv14, pw16, sg18}, provide robust evidence to strongly support the hidden BLRs in type-2 
AGN. Certainly, there is one precious subclass of true type-2 AGN: type-2 AGN without hidden BLRs. 
More recent arguments on the very existence of true type-2 AGN can be found in \citet{ba19, zz21}. We 
mainly focus on properties of NLRs sizes of AGN, therefore, we do not discuss true type-2 
AGN in the manuscript.

	Basic physical sizes of NLRs in type-2 AGN and of BLRs in type-1 AGN have been well known. 
Sizes of BLRs ($R_{BLRs}$ distances between BLRs to central black hole (BH)) can be well determined 
by the well-known reverberation mapping technique \citep{bm82}, through long-term variabilities of 
broad emission lines and continuum emissions. \citet{pf04} have reported the measured sizes 
of BLRs in the sample of nearby broad line AGN (BLAGN) in the AGNWATCH project\footnote{
\url{http://www.astronomy.ohio-state.edu/~agnwatch}} (see dozens of published papers listed in the project). 
\citet{kas00, km05} have reported the measured sizes of BLRs of nearby 17 quasars\footnote{
\url{http://wise-obs.tau.ac.il/~shai/PG/}}. \citet{bd10, ba15, wp18} have reported the measured sizes 
of BLRs in the sample of nearby Seyfert galaxies in the LAMP project\footnote{
\url{https://www.physics.uci.edu/~barth/lamp.html}}. \citet{wa04, du16} have reported the 
measured sizes of BLRs in the sample of nearby Seyferts with high accretion rates. \citet{sh15, gt17} 
have reported the measured sizes of BLRs in the sample of SDSS QSOs\footnote{
\url{https://www.sdss.org/dr15/algorithms/ancillary/reverberation-mapping/}}. The reverberation 
mapping technique determined sizes of BLRs leads to the well accepted empirical relation between 
$R_{BLRs}$ and continuum luminosity of broad line AGN discussed and shown in \citet{kas00, km05, bd13}.

    Unlike the measurements of BLRs sizes in type-1 AGN, longer extended structures 
of NLRs indicate reverberation mapping technique can not be applied to measure NLRs sizes ($R_{NLRs}$) 
due to no variabilities in narrow emission lines (expected variability timescale around hundreds of 
years). However, in some nearby type-2 AGN or nearby high luminosity type-2 QSOs, the NLRs sizes have 
been well determined by properties of spatial resolved \o3~ emission images, such as the measured NLRs 
sizes in Seyfert 2 galaxy NGC1386\ in \citet{bg06a}, in the small sample of in Seyfert-2 
galaxies in \citet{bg06b}, in the sample of 15 obscured QSOs in \citet{gh11}, in the radio quiet type-2 
QSOs in \citet{liu13, lu13b}, in the luminous obscured QSOs in \citet{ha13, ha14}, in the obscured QSOs 
in \citet{fk18}, etc. Then, based on the measured sizes of NLRs, there is one empirical R-L relation 
between NLRs sizes and \o3~ luminosity ($L_{[O~\textsc{iii}]}$) (in the manuscript, the R-L relation 
is for the relation on NLRs sizes, unless with specific statements) discussed and shown in \citet{liu13, 
ha13, ha14, fk18, dz18},
\begin{equation}
\begin{split}
\log(\frac{R_{NLRs}}{\rm pc})=&(0.250\pm0.018)\times\log(\frac{L_{[O~\textsc{iii}]}}{\rm 10^{42}erg/s})\\ 
	& + (3.746\pm0.028)
\end{split}
\end{equation},
especially for AGN with 8$\mu$m luminosity smaller than $10^{45}{\rm erg/s}$ as discussed in \citet{ha13, 
dz18}. In order to explain the R-L relation of NLRs, \citet{liu13} have proposed a model of 
\o3~ emission clumpy nebulae in which clouds that produce line emission transition from being 
ionization-bounded at small distances from central power source to being matter-bounded in the outer parts 
of the clumpy nebulae. Similar model was also discussed in \citet{gz11}. More recently, \citet{dz18} 
have re-constructed the R-L relation of NLRs by the proposed model that the NLRs as a collection of clouds 
in pressure equilibrium with the ionizing radiation.

	Actually, NLRs sizes in type-1 AGN are hardly to be measured mainly due to inclination effects, 
although \citet{bf02, sc03a, sc03b} have reported NLRs sizes in several nearby broad line AGN. 
However, the measured NLRs sizes of the type-1 AGN in the literature, such as the results in \citet{sc03b}, 
are about 1-2 magnitudes smaller than the expected results from the empirical relation between NLRs sizes 
and \o3~ luminosity in the type-2 AGN in \citet{liu13}, which could be mainly due to inclination effects 
and/or quite shallow observations. Therefore, NLRs sizes of type-1 AGN in \citet{sc03b} are not considered 
in the manuscript. However, there is one subclass of type-1 AGN, the broad line AGN with double-peaked 
broad low-ionization emission lines (called as double-peaked BLAGN) of which inclination angles can be 
reasonably estimated, accepted accretion disk origin of the double-peaked broad emission lines as well 
discussed in \citet{ch89, el95, sn03, st03, fe08, sb17}, etc. In the manuscript, considering the fixed 
diameter of SDSS fibers, upper limits of NLRs sizes in low redshift double-peaked BLAGN in SDSS can be 
well estimated, which will be applied to test the R-L empirical relation of NLRs in type-1 AGN. The tests 
can provide further clues on whether there are difference of spatial distances of NLRs between type-1 AGN 
and type-2 AGN, to challenge or to support the Unified model of AGN, which is the main objective of the 
manuscript. Section 2 presents our hypotheses on estimating sizes of NLRs of double-peaked BLAGN. Section 
3 shows our main sample of double-peaked BLAGN. Section 4 shows our main results and necessary discussions. 
Section 5 gives our final summaries and conclusions. We have adopted the cosmological parameters 
$H_{0}=70{\rm km\cdot s}^{-1}{\rm Mpc}^{-1}$, $\Omega_{\Lambda}=0.7$ and $\Omega_{\rm m}=0.3$ in the 
manuscript.

\begin{figure*}
\centering\includegraphics[width = 18cm,height=20cm]{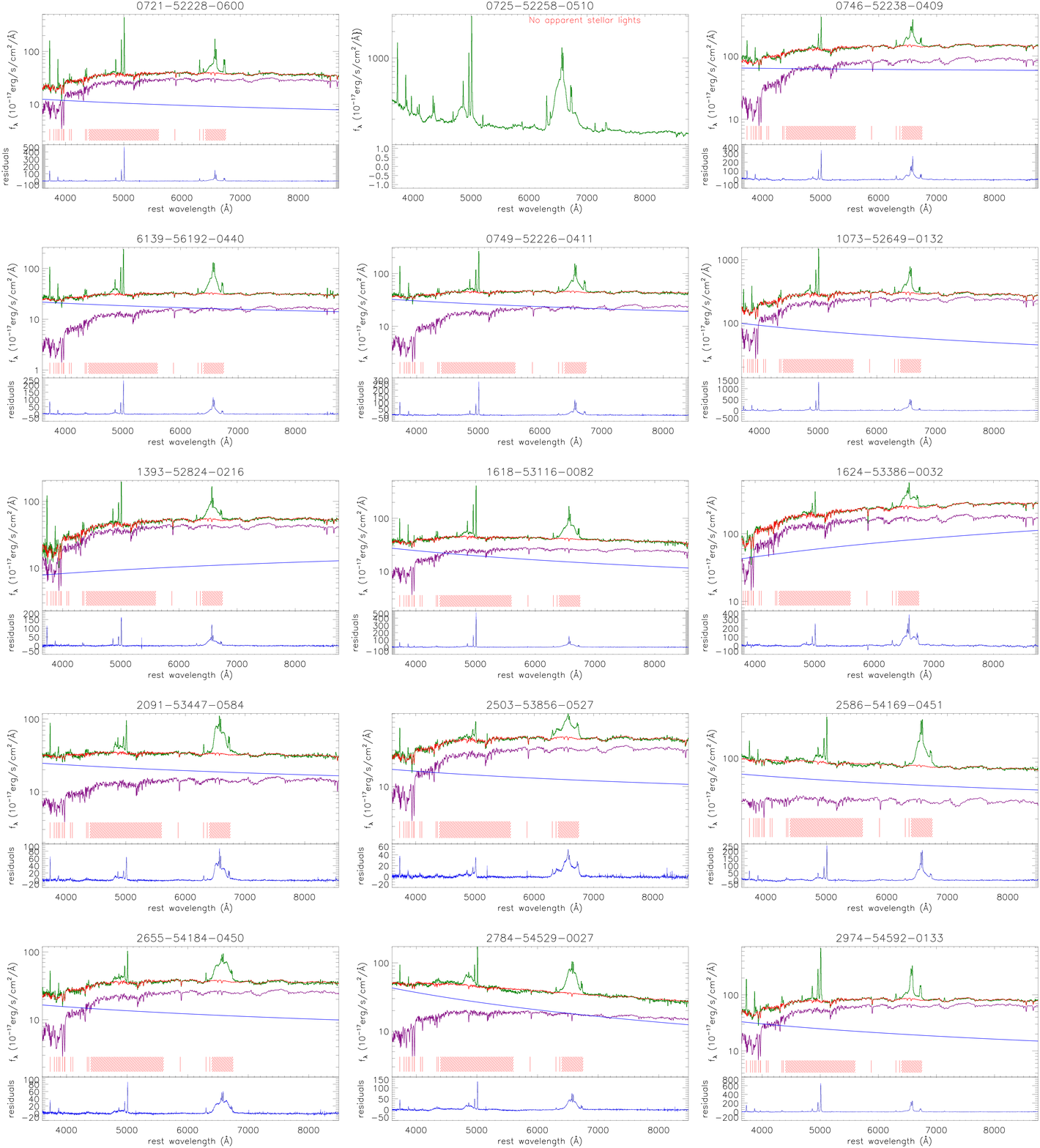}
\caption{SDSS spectra of the 38 low redshift double-peaked BLAGN (marked with information of 
PLATE-MJD-FIBERID, and listed in the order shown in Table~1) and the best descriptions to the stellar 
lights by the SSP method. In each panel (except the ones without apparent stellar lights), solid dark green 
line shows the SDSS spectrum, solid red line shows the best descriptions to the SDSS spectrum with the 
emission lines being masked out, solid purple line shows the SSP method determined stellar lights, and solid 
blue line shows the determined power law AGN continuum emissions. In each panel, from left to right, the 
vertical red lines mark the following emission features masked out, including \oii, H$\theta$, H$\eta$, 
[Ne~{\sc iii}]$\lambda3869$\AA, He~{\sc i}$\lambda3891$\AA, Calcium K line, [Ne~{\sc iii}]$\lambda3968$\AA, 
Calcium H line, [S~{\sc ii}]$\lambda4070$\AA, H$\delta$, H$\gamma$, [O~{\sc iii}]$\lambda4364$\AA, 
He~{\sc i}$\lambda5877$\AA\ and [O~{\sc i}]$\lambda6300,6363$\AA, respectively, and the area filled by 
red lines around 5000\AA\ shows the region masked out including the emission features of probable optical 
Fe~{\sc ii} lines, broad and narrow H$\beta$ and [O~{\sc iii}] doublet, and the area filled by red lines 
around 6550\AA\ shows the region masked out including the emission features of broad and narrow H$\alpha$, 
[N~{\sc ii}] and [S~{\sc ii}] doublets.Bottom region of each panel shows the residuals 
calculated by SDSS spectrum minus sum of the stellar lights and the power law continuum emissions, for 
the object with apparent stellar lights. {\bf In order to show clearly spectral features and features 
of stellar lights, the Y-axis is in logarithmic coordinate for the plots in top region of each panel.}
}
\label{spec}
\end{figure*}

\setcounter{figure}{0}   

\begin{figure*}
\centering\includegraphics[width = 18cm,height=22cm]{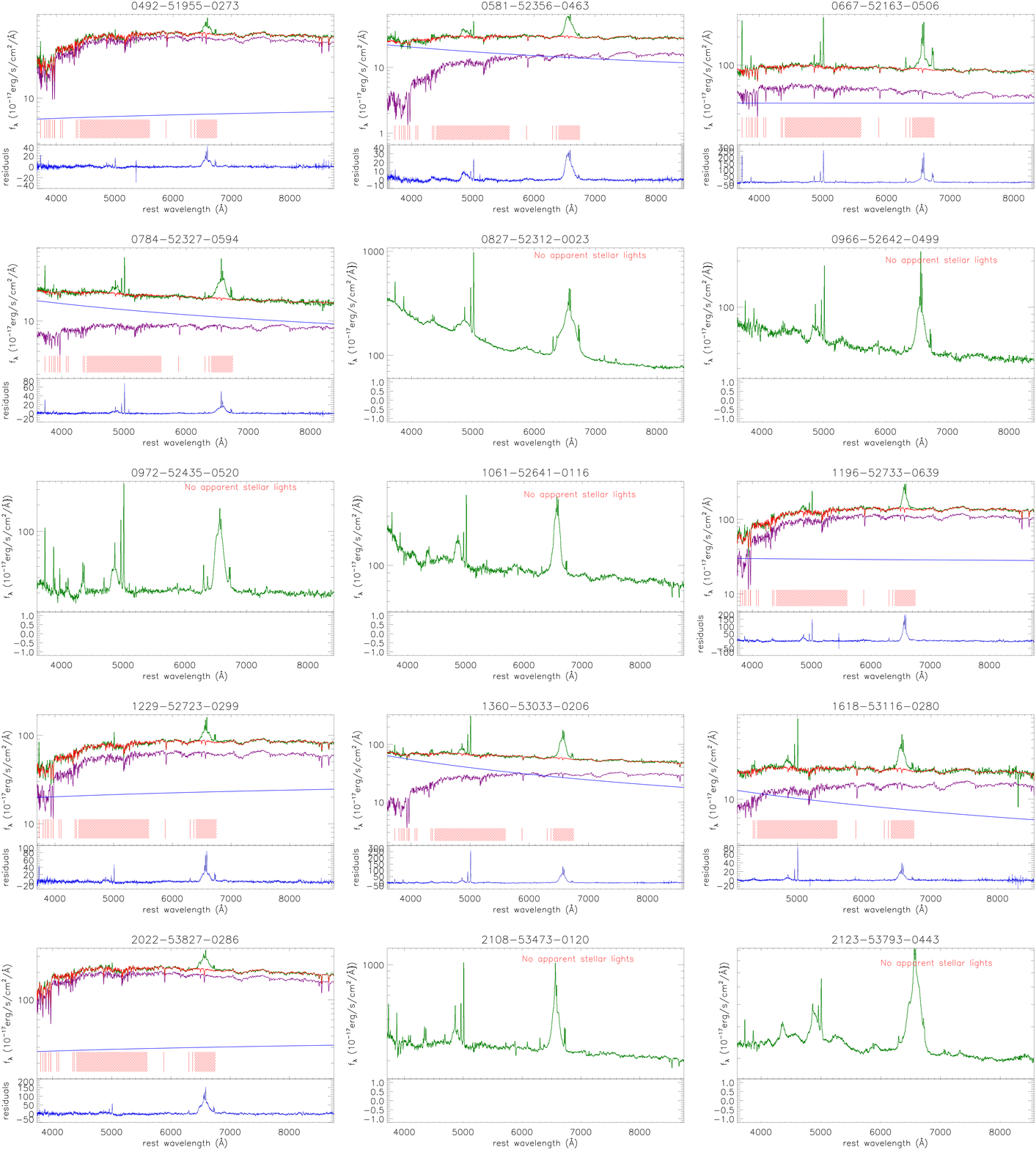}
\caption{--continued.}
\end{figure*}

\setcounter{figure}{0}

\begin{figure*}[htp]
\centering\includegraphics[width = 18cm,height=12cm]{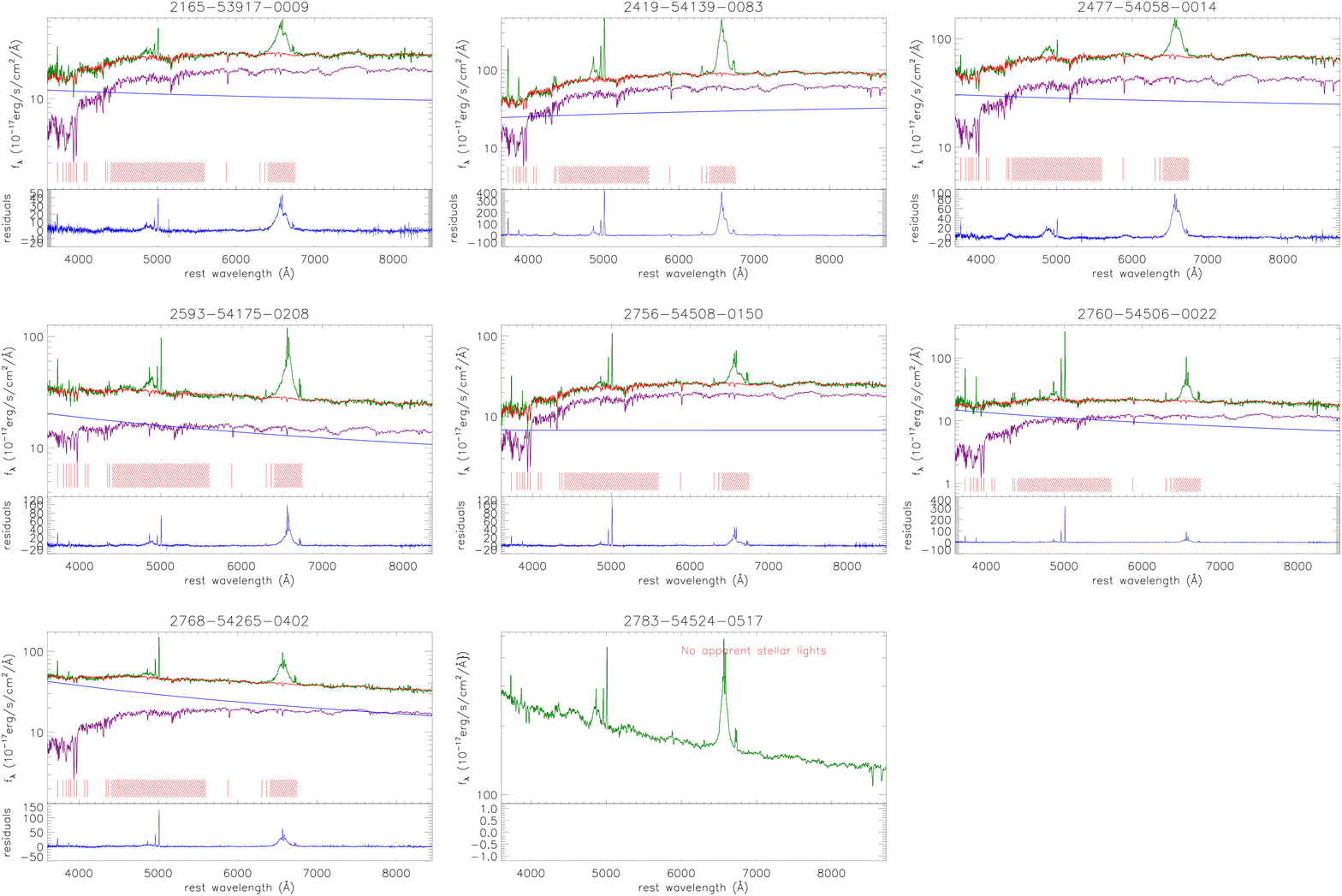}
\caption{--continued. 
}
\end{figure*}

\section{Main Hypotheses on estimating upper limits of sizes of NLRs of double-peaked BLAGN}

	In order to explain the double-peaked broad emission lines of AGN, besides the accretion disk 
origin \citep{ch89, el95, sb17}, the theoretical binary black hole model (BBH model) \citep{bb80, dm05, 
pl17, sb21} has been well proposed, such as the BBH model once applied to describe variability properties 
of the double-peaked broad Balmer lines of 3C390.3\ in \citet{gm96} and theoretical BBH model applied 
to generate double-peaked broad lines in \citet{sl10}, etc. Once accepted the BBH model to explain the 
double-peaked broad emission lines, quasi-periodic oscillations (QPOs) could be expected, such as the 
well discussed in \citet{gd15a, gm15, zb16, lw21, zh22} with QPOs signs applied to detect candidates for BBH 
systems. However, through the study of long-term variabilities of double-peaked BLAGN in \citet{era97, 
sh01, sn03, lew10, zh13, zh17}, quite few QPOs can be commonly detected in long term variabilities of 
double-peaked BLAGN. Moreover, among the 38 collected low-redshift double-peaked BLAGN in our final 
sample in the following section, 37 double-peaked BLAGN have their long-term light curves collected from 
CSS (Catalina Sky Survey\footnote{\url{http://nesssi.cacr.caltech.edu/DataRelease/}}) \citep{dd09}. 
Long-term variability analysis of our sources does not show any QPOs signs as shown in section 4. Therefore, 
in the manuscript, the accretion disk origin has been well accepted to explain the double-peaked broad 
emission lines, and no further discussions are shown on the BBH model.

	There are different kinds of relativistic accretion disk models in the literature, such as 
the circular accretion disk model in \citet{ch89}, the improved elliptical accretion disk model in 
\citet{el95}, the model of circular disk with spiral arms in \citet{sn03}, a warped accretion disk 
model in \citet{hb00}, and a stochastically perturbed accretion disk model in \citet{fe08}, etc. 
Here, the elliptical accretion disk model (without contributions of subtle structures) well discussed 
in \citet{el95} is preferred, because the model can be applied to explain almost all observational 
double-peaked features of the double-peaked BLAGN in our sample. There are seven model parameters 
in the elliptical accretion disk model, inner boundary $r_0$ and out boundary $r_1$ in the units of 
$R_G$ (Schwarzschild radius), inclination angle $i$ of disk-like BLRs, eccentricity $e$, orientation 
angle $\phi_0$ of elliptical rings, local broadening velocity $\sigma_L$, line emissivity slope $q$ 
($f_r~\propto~r^{-q}$). Meanwhile, we have also applied the very familiar elliptical accretion disk 
model, see our studies on double-peaked lines in \citet{zh05, zh13a, zh15, zh19}. More detailed 
descriptions on the applied elliptical accretion disk model can be found in \citet{el95, sn03, st03}, 
and there are no further discussions on the elliptical accretion disk model in the manuscript. Then, 
the elliptical accretion disk model can be well applied to describe double-peaked broad emission 
lines, leading to the well determined inclination angle $\sin(i)$ of central accretion disk.

	Considering the SDSS fiber diameter of 3 arcseconds (2 arcseconds for eBOSS, the Extended 
Baryon Oscillation Spectroscopic Survey\footnote{\url{https://www.sdss.org/surveys/eboss}} (detailed 
descriptions can be found in the web\footnote{\url{https://www.sdss.org/dr16/spectro/spectro_basics/}}) 
as the projected space distance $D_s$ in units of $pc$, and considering the inclination angle $\sin(i)$ 
determined by the elliptical accretion disk model, the upper limits of sizes of the \o3~ emission regions, as the 
upper limits of NLRs sizes $R_{NLRs, u}$, should be estimated as  
\begin{equation}
R_{NLRs, u} = \frac{D_s}{sin(i)}
\end{equation}.

	Before the end of the section, it is interesting to check whether the equation above can be 
reasonably applied to estimate upper limits of NLRs sizes of AGN. For the 14 obscured quasars well 
discussed in \citet{liu13}, their upper limits of NLRs sizes can be estimated as SDSS fiber radius 
of 1.5 arcseconds. Here, for the obscured quasars (type-2 quasar), we simply accepted $sin(i)\sim1$. 
Due to the high redshift around 0.5 of the 14 obscured quasars, their upper limits of NLRs sizes are 
about $2.06\times10^4pc$, quite larger than their measured NLRs sizes in \citet{liu13}, indicating 
the method to estimate upper limits of NLRs sizes are reasonable to some extent.

\section{Sample of the double-peaked BLAGN} 

	In the manuscript, we mainly consider the low redshift double-peaked BLAGN with redshift 
smaller than 0.1 by the following main reason. For redshift at 0.1 (distance about 460.3Mpc), the 
corresponding space distance of the SDSS fiber radius (1.5 arcseconds) is about $3350{\rm pc}$ 
($10^{3.53}{\rm pc}$), similar as the mean value about $10^{3.5}{\rm pc}$ of the measured NLRs sizes 
of type-2 QSOs in \citet{liu13, ha13, fk18}. For objects with redshift quite larger than 0.1, the 
corresponding space distance by the SDSS fibers should be clearly larger than the expected sizes of 
NLRs, indicating that it is meaningless to check properties of upper limits of NLRs sizes of type-1 
objects with redshift quite larger than 0.1.

	We collect double-peaked BLAGN with redshift smaller than 0.1 as follows. On the one hand, there 
are 4 double-peaked BLAGN with redshift smaller than 0.1\ in the sample of double-peaked BLAGN listed in 
\citet{st03}, SDSS 0721-52228-0600 (SDSS PLATE-MJD-FIBERID), SDSS 0725-52258-0510, SDSS 0746-52238-0409 
and SDSS 6139-56192-0440 (spectrum from eBOSS with fiber radius 1arcseconds). On the other hand, the 
other 11 appropriate double-peaked BLAGN are collected by the following two steps. First, all the 1752 
SDSS QSOs with signal-to-noise larger than 10 and redshift smaller than 0.1 are firstly collected from 
SDSS DR16 (Data Release 16 \citep{ap20}) by the following SQL query\footnote{
\url{http://skyserver.sdss.org/dr16/en/tools/search/sql.aspx}}
\begin{lstlisting}
SELECT
   plate, fiberid, mjd
FROM SpecObjall
WHERE
   class = 'QSO' and z < 0.1
   and zwarning = 0 and snmedian > 10
\end{lstlisting}.
Second, emission lines of all the 1752 SDSS QSOs are measured after host contributions are determined, 
as what are discussed in the following section. And then, we mainly check the QSOs with the broad 
H$\alpha$ described by more than 2 broad Gaussian components one by one by eyes, and collect 11 
double-peaked BLAGN based on the criterion that there is {\bf at least one apparent hump} included in 
the red-side or blue-side of broad H$\alpha$. Unfortunately, it is hard to provide a standard criterion 
which can be well described by a formula or formulas to collect candidates of AGN with double-peaked 
broad emission lines, therefore, the collected double-peaked BLAGN are identified by eyes.

	Besides the double-peaked BLAGN collected from SDSS quasars in DR16, there is a sample of 
candidates of double-peaked BLAGN collected from SDSS main galaxies and reported in \citet{liu19}. 
Then, based on the same criteria of $z~<~0.1$ and signal-to-noise of SDSS spectrum larger than 10, 
there are 106 candidates of double-peaked BLAGN collected from the sample of \citet{liu19} with the 
flag MULTI\_PEAK=2. Furthermore, two additional criteria are accepted to collect double-peaked 
BLAGN from the sample of \citet{liu19}. On the one hand, there should be apparent narrow Balmer 
emission lines detected in the SDSS spectra, in order to correct reddening effects on observed 
[O~{\sc iii}] emission intensities. On the other hand, there should be apparent both broad H$\alpha$ 
and broad H$\beta$ detected in the SDSS spectra, in order to ignore seriously obscurations on central 
BLRs, and in order to confirm the collected double-peaked BLAGN from SDSS main galaxies are also 
type-1 AGN (not type-1.5 nor type-1.9 AGN) totally similar as the double-peaked BLAGN (type-1 AGN) 
collected from SDSS quasars. Based on the two additional criteria, 23 double-peaked BLAGN 
are finally collected from the sample of \citet{liu19}. In the following section, two examples 
are shown on an AGN with double-peaked broad H$\alpha$ but no apparent narrow Balmer emission 
lines, and on an AGN with only double-peaked broad H$\alpha$ but no apparent double-peaked broad 
H$\beta$, from the sample of \citet{liu19}, after subtractions of host galaxy contributions from 
the SDSS spectra. Finally, there are 38 double-peaked BLAGN with redshift smaller than 0.1 included 
in our sample, 15 double-peaked BLAGN collected from SDSS quasars and 23 double-peaked BLAGN 
collected from the sample of \citet{liu19}. The basic information of the 38 double-peaked BLAGN are 
listed in Table~1, including the PLATE-MJD-FIBERID, redshift, etc.

\begin{figure*}[htp]
\centering\includegraphics[width = 18cm,height=20cm]{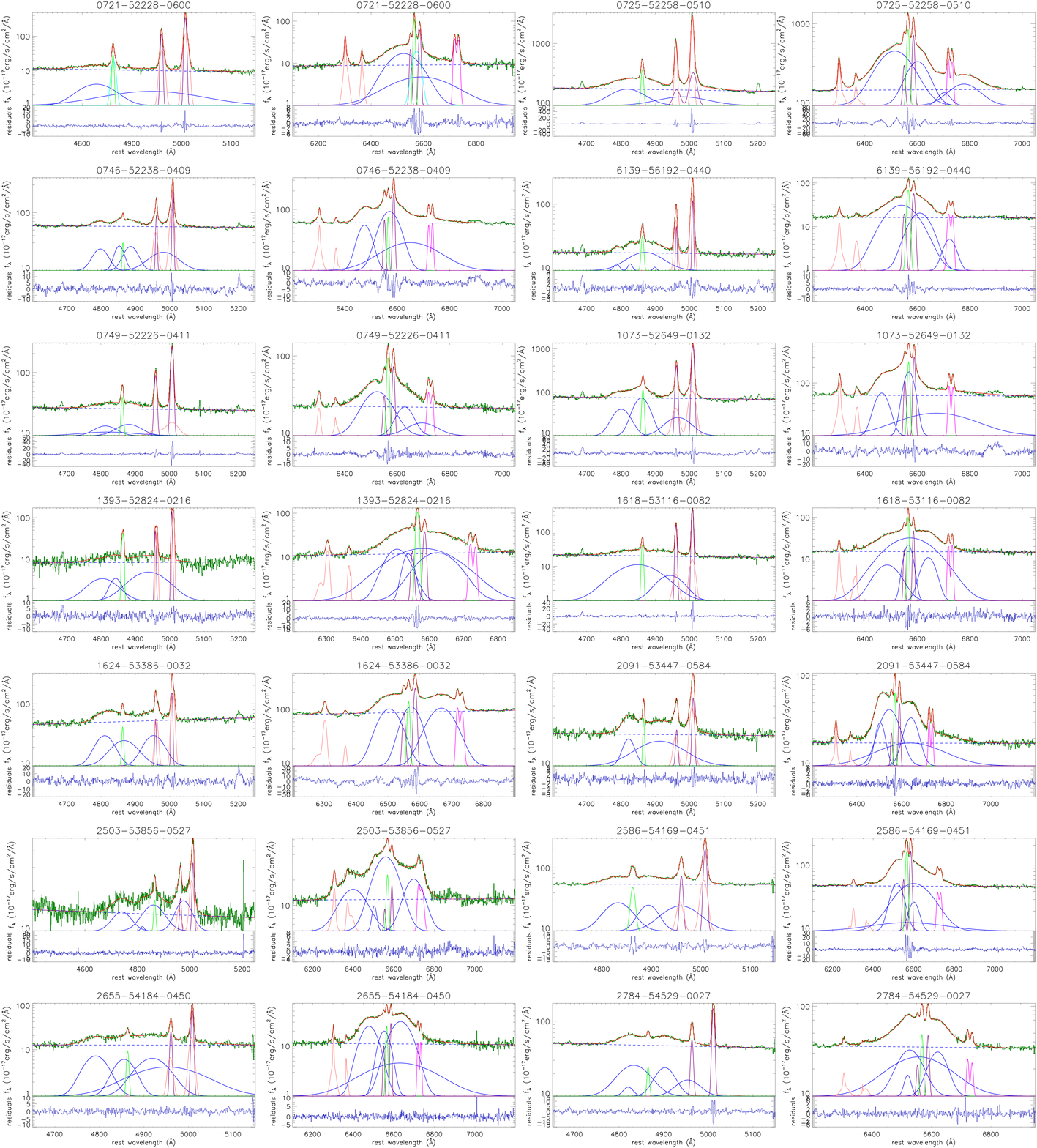}
\caption{The best fitting results to the emission lines around H$\beta$ (panels in the 
first and the third column) and around H$\alpha$ (panels in the second and the fourth 
column) by multiple Gaussian functions. In each panel, solid line in dark green shows the line 
spectrum after subtractions of the contributions of stellar lights, solid red line shows the best 
fitting results by multiple Gaussian components plus a power law component, solid blue lines show 
the determined broad Gaussian components in the Balmer line, solid green line and solid cyan line (if 
there was) show the determined narrow and extended components in the Balmer line, and dashed blue 
line shows the determined power law component. In each panel in the first and the third 
column, solid pink line and solid purple line show the determined narrow and extended components 
in [O~{\sc iii}] doublet. In each panel in the second and the fourth column, solid 
lines in pink show the determined [O~{\sc i}] doublet including both narrow and extended components, 
solid purple lines show the determined [N~{\sc ii}] doublet, solid lines in magenta show the 
determined [S~{\sc ii}] doublet. In the panels for best fitting results around H$\beta$ 
of SDSS 0966-52642-0499, SDSS 2123-53793-0443 and SDSS 2783-54524-0517, dashed cyan lines show the 
determined optical Fe~{\sc ii} emission features. Bottom region of each panel shows the 
residuals calculated by line spectrum minus the best fitting results. {\bf In order to show clearly 
spectral broad emission line features, the Y-axis is in logarithmic coordinate for the plots in top 
region of each panel.}
}
\label{line}
\end{figure*}

\setcounter{figure}{1}

\begin{figure*}[htp]
\centering\includegraphics[width = 18cm,height=22cm]{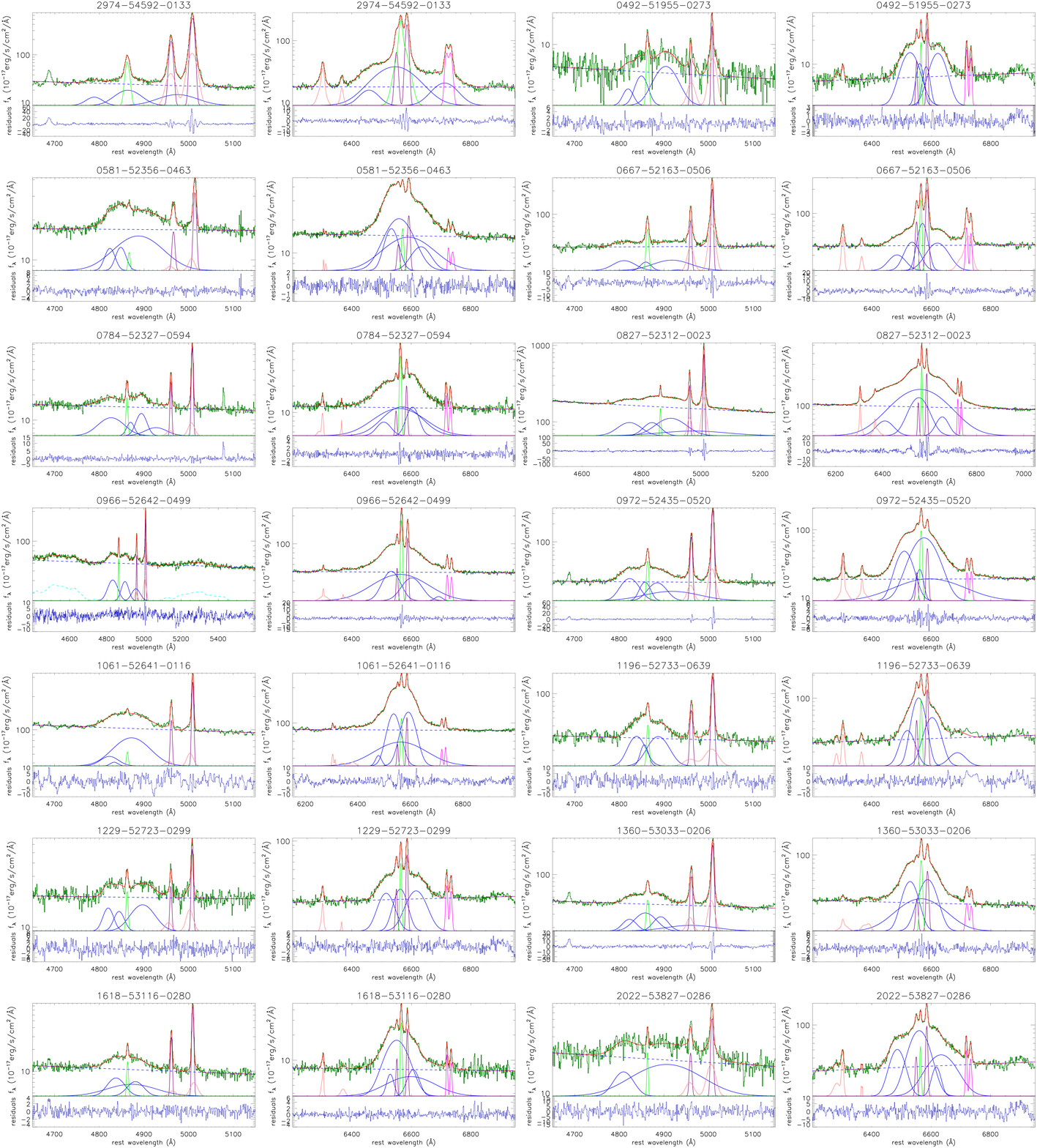}
\caption{--continued.
}
\end{figure*}

\setcounter{figure}{1}

\begin{figure*}[htp]
\centering\includegraphics[width = 18cm,height=15cm]{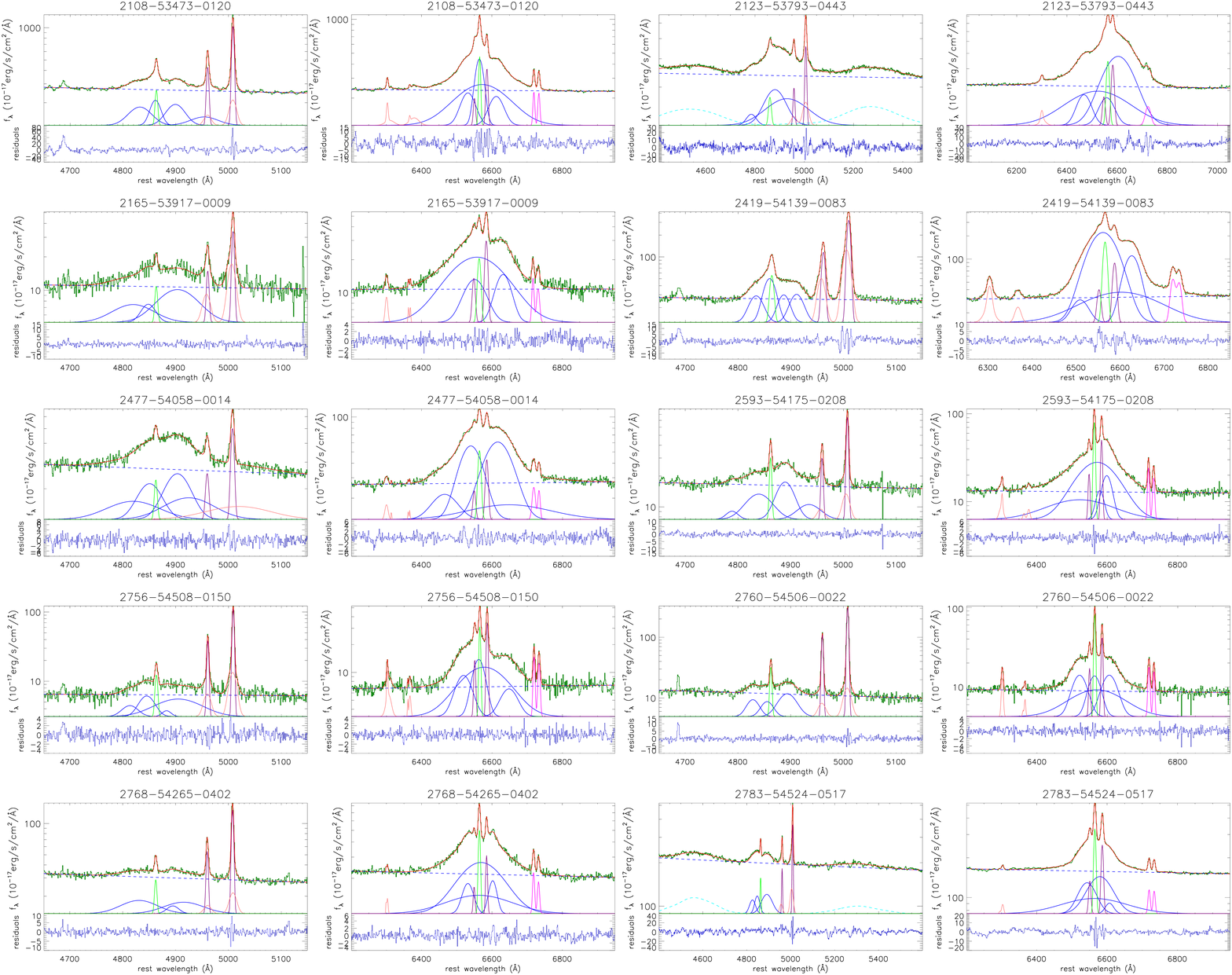}
\caption{--continued.}
\end{figure*}

\begin{figure*}[htp]
\centering\includegraphics[width = 18cm,height=8cm]{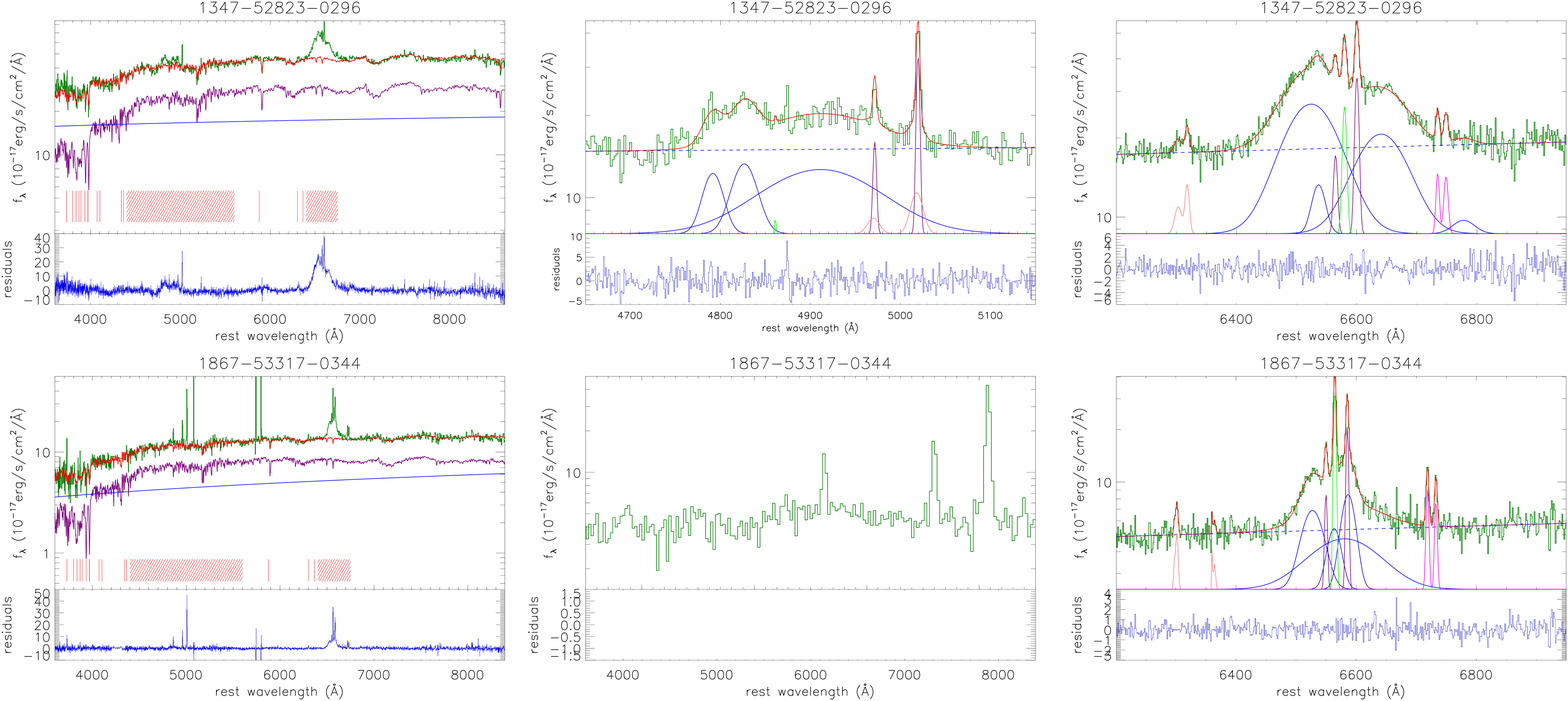}
\caption{Two examples on the BLAGN with double-peaked broad H$\alpha$ identified in \citet{liu19}, 
but without apparent narrow H$\beta$ in SDSS 1347-52823-0296 (top panels), without apparent broad H$\beta$ in 
SDSS 1867-53317-0344 (bottom panels). Each left panel shows the SDSS spectrum and the SSP method determined 
host galaxy contributions, symbols and line style have the same meanings as those in Figure~\ref{spec}. 
Middle and right panels show the best descriptions to the emission lines around H$\beta$ and around H$\alpha$, 
symbols and line style have the same meanings as those in Figure~\ref{line}. Here, due to lack of apparent broad 
H$\beta$, there are no fitting results shown in bottom middle panel.}
\label{l2ine}
\end{figure*}

\section{Main procedures, Main results and Main discussions}
   
	For the 38 double-peaked BLAGN, the SDSS spectra are shown in Figure~\ref{spec}. It is clear that 
there are apparent stellar lights included in the spectra of 30 double-peaked BLAGN. The stellar lights 
should be firstly determined, in order to find more accurate line profiles of emission lines, especially 
of the narrow emission lines. In the manuscript, the well accepted SSP (Simple Stellar Population) method 
is applied. Detailed discussions and descriptions on the SSP method can be found in \citet{bc03, ka03, cm05}, 
here we do not show them any more. Similar as what we have previously done in \citet{zh14, zh16, zh19, zh21, 
zh21b} to describe the stellar lights included in SDSS spectra of broad line AGN, we here exploit a power 
law component $\alpha\times\lambda^\beta$ applied to describe the AGN continuum emissions and the 39 simple 
stellar population templates from the \citet{bc03} which include the population age from 5Myr to 12Gyr, 
with three solar metallicities (Z = 0.008, 0.05, 0.02). Then through the Levenberg-Marquardt least-squares 
minimization method applied to the SDSS spectra with both narrow and broad emission lines being masked out, 
host galaxy contributions and the intrinsic AGN power law continuum components in the SDSS spectra can be 
clearly determined. Here, when the SSP method is applied, the emission lines\footnote{
\url{http://classic.sdss.org/dr1/algorithms/reflines.dat}} are being masked out by line width about 
400${\rm km/s}$ at zero intensity, and the broad H$\alpha$, H$\beta$ and H$\gamma$ are being masked out by 
line width about 3000${\rm km/s}$ at zero intensity, and the probable optical Fe~{\sc ii} emission lines 
\citep{kp10} are being masked out with rest wavelength range from 4400\AA~ to 5600\AA. Meanwhile, when the 
model functions are applied, there are 42 model parameters, 39 strengthen factors with zero as the starting 
values for the 39 SSPs, the broadening velocity with ${\rm 100km~s^{-1}}$ as the starting value, $\alpha$ 
and $\beta$ for the power law component with zero as the starting values. The SSP method determined stellar 
lights and corresponding residuals are clearly shown in Figure~\ref{spec} for the 30 double-peaked BLAGN 
with apparent stellar lights. Here, the residuals shown in Figure~\ref{spec} are calculated by SDSS spectrum 
minus sum of the stellar lights and the power law continuum emissions. The values of $\chi^2$ (summed squared 
residuals divided by degree of freedom) are listed in Table~1 for the best fitting results to the SDSS spectra 
with emission lines being masked out.

	Before subtracting the stellar lights from the SDSS spectra, one point is noted. Among the 30 
double-peaked BLAGN of which SDSS spectra include apparent contributions of stellar lights, there are six 
objects, SDSS 1393-52824-0216, SDSS 1624-53386-0032, SDSS 0492-51955-0273, SDSS 1229-52723-0299, 
SDSS 2022-53827-0286 and SDSS 2419-54139-0083, of which intrinsic AGN continuum emissions are determined 
and described by red power law functions ($\beta~\ge~0$) not by commonly blue power law functions ($\beta~<~0$). 
The red power law AGN continuum emissions are probably due to intrinsic reddening effects of host galaxies 
and/or central high density clouds. The intrinsic reddening could have effects on strengths of broad emission 
lines but have few effects on line profiles of broad emission lines. Whereas, the main consider parameter of 
inclination angle is determined through properties of double-peaked line profiles. Therefore, although in 
the procedure above to determine contributions of stellar lights without considerations of intrinsic 
reddening effects, there are few effects on our final results on the upper limits of NLRs sizes of the 
collected low redshift double-peaked BLAGN.

	After subtractions of the contributions of stellar lights, the emission lines 
are modeled by multiple Gaussian functions, in order to obtain more accurate line parameters (or line 
profiles) of both narrow and broad emission lines. Similar as what we have recently done in \citet{zh21}, 
the following model functions are applied to describe the emission lines around H$\alpha$ within rest 
wavelength range from 6250\AA~ to 6850\AA, four broad Gaussian functions (second moment larger than 
400${\rm km/s}$) applied to describe the broad H$\alpha$, one narrow Gaussian function (second moment 
smaller than 400${\rm km/s}$) plus one broad Gaussian function applied to describe the narrow and 
extended (if there was) components of narrow H$\alpha$\footnote{Similar as the extended components in 
[O~{\sc iii}] doublet, there are also extended components in the narrow Balmer lines, such as the shown 
results in SDSS 0721-52228-0600 (PLATE-MJD-FIBERID). Here, we do not discuss the physical properties 
and origin of the extended narrow Balmer components, but the applied extended components can lead to 
better descriptions to the emissions lines shown in Figure~\ref{line}.}, two narrow Gaussian functions 
applied to describe the [N~{\sc ii}] doublet, two narrow Gaussian functions applied to describe the 
[S~{\sc ii}] doublet, and two narrow plus two broad Gaussian functions applied to describe the [O~{\sc i}] 
doublet\footnote{Similar as the extended components in [O~{\sc iii}] doublet, there are also extended 
components in the [O~{\sc i}] doublet, such as the shown results in SDSS 1073-52649-0132 (PLATE-MJD-FIBERID). 
Here, we do not discuss the physical properties and origin of the extended [O~{\sc i}] components, 
but the applied extended components can lead to better descriptions to the emissions lines around 
H$\alpha$ shown in Figure~\ref{line}.} and a power law component $\alpha_1\times\lambda^\beta_1$ applied 
to describe the AGN continuum emissions underneath the emission lines around H$\alpha$. Similarly, 
the following model functions are applied to describe the emission lines around H$\beta$ within rest 
wavelength range from 4650\AA~ to 5250\AA, four broad Gaussian functions applied to describe the broad 
H$\beta$, one narrow plus one broad Gaussian functions applied to describe the narrow and extended 
components of narrow H$\beta$, two narrow plus two broad Gaussian functions applied to describe the 
[O~{\sc iii}] doublet and a power law component $\alpha_2\times\lambda^\beta_2$ applied to describe 
the intrinsic AGN continuum emissions underneath the emission lines around H$\beta$. 

     When the model functions are applied to describe emission lines around H$\alpha$, there are 44 
model parameters, 12 model parameters from the 4 broad Gaussian components (three model parameters 
for each broad Gaussian component, central wavelength $\lambda_{0b,~i}$, second moment $\sigma_{b,~i}$ 
and line flux $f_{b,~i}$) for the broad H$\alpha$, 21 model parameters from the 7 narrow Gaussian 
components (three model parameters for each Gaussian component, central wavelength $\lambda_{0n,~j}$, 
second moment $\sigma_{n,~j}$ and line flux $f_{n,~j}$), 9 model parameters from the three extended 
components of narrow H$\alpha$ and [O~{\sc i}] doublet ($\lambda_{0e,~k}$, $\sigma_{e,~k}$ and 
$f_{e,~k}$) and 2 model parameters $\alpha_1$ and $\beta_1$ for the power law component. The 
restrictions on the 44 model parameters are as follows. First, all broad and narrow Gaussian components 
have line fluxes not smaller than zero $f_{b,~i}~\ge~0~(i=1,~\dots,~4)$, $f_{n,~j}~\ge~0~(j=1,~\dots,~7)$ 
and $f_{e,~k}~\ge~0~(k=1,~\dots,~3)$, and have zero as the starting values of $f_{b,~i}$, $f_{n~,j}$ 
and $f_{e,~k}$. Second, the 7 narrow Gaussian components have second moments smaller than 
400${\rm km/s}$,  $0~<~\sigma_{n,~j}~\le~400{\rm km/s}~(j=1,~\dots,~7)$, and have 100${\rm km/s}$ 
as the starting values. Third, the 4 broad Gaussian components in broad Balmer lines have second 
moments larger than 400${\rm km/s}$,  $\sigma_{b,~i}~\ge~400{\rm km/s}~(i=1,~\dots,~4)$, and have 
1000${\rm km/s}$ as the starting values. Fourth, the 3 broad Gaussian components for the extended 
emission components in narrow emission lines have second moments larger than those of the corresponding 
narrow components, but have 600${\rm km/s}$ as the starting values. Fifth, the 7 narrow Gaussian 
components have central wavelengths in units of \AA~ to be within the range 
$\lambda_{0T,~j}-15\textsc{\AA}~\le~\lambda_{0n,~j}~\le~\lambda_{0T,~j}+15\textsc{\AA}$ with 
$\lambda_{0T,~j}~(j=1,~\dots,~7)$ as the theoretical vacuum rest wavelength of the emission component, 
and have $\lambda_{0T,~j}$ as the starting values. Sixth, the 3 broad Gaussian functions for the 
extended emission components have central wavelengths in units of \AA~ to be within the range 
$\lambda_{0T,~k}-30\textsc{\AA}~\le~\lambda_{0e,~k}~\le~\lambda_{0T,~k}+30\textsc{\AA}$ with 
$\lambda_{0T,~k}~(k=1,~\dots,~3)$ as the theoretical vacuum rest wavelength of the emission component, 
and have $\lambda_{0T,~k}$ as the starting values. Seventh, the 4 broad Gaussian components for broad 
H$\alpha$ have central wavelengths in units of \AA~ to be within the range 
$6450\textsc{\AA}~\le~\lambda_{0b,~i}~\le~6650\textsc{\AA}$, and have 6490\AA, 6520\AA, 6580\AA, 
6620\AA~ as the starting values. Eighth, $\alpha_1$ and $\beta_1$ are restricted by $\alpha_1~\ge~0$ 
and $-5~\le~\beta_1~\le~1$. Ninth, the flux ratio of [N~{\sc ii}] doublet is fixed to the theoretical 
values of 3. Tenth, the Gaussian components from each doublet have the same redshift and the same 
second moment. 

	Similar restrictions are applied to the 32 model parameters, when the model functions are 
applied to describe the emission lines around H$\beta$. There are 12 model parameters from the 4 
broad Gaussian components ($\lambda_{0b,~i}$, $\sigma_{b,~i}$, $f_{b,~i}$, $i=1,~\dots,~4$) for 
the broad H$\beta$, 9 model parameters from the 3 narrow Gaussian components ($\lambda_{0n,~j}$, 
$\sigma_{n,~j}$, $f_{n,~j}$), 9 model parameters from the three extended components of narrow H$\beta$ 
and [O~{\sc iii}] doublet ($\lambda_{0e,~k}$, $\sigma_{e,~k}$ and $f_{e,~k}$) and 2 model parameters 
$\alpha_2$ and $\beta_2$ for the power law. For the parameters of the narrow Gaussian components 
($j=1,~\dots,~3$), we have $f_{n,~j}~\ge~0$, $0~<~\sigma_{n,~j}~\le~400{\rm km/s}$, 
$\lambda_{0T,~j}-15\textsc{\AA}~\le~\lambda_{0n,~j}~\le~\lambda_{0T,~j}+15\textsc{\AA}$, and have 
zero as staring values of $f_{n,~j}$, 100${\rm km/s}$ as starting values of $\sigma_{n,~j}$ and 
$\lambda_{0T,~j}$ as starting values of $\lambda_{0n,~j}$. For the parameters of the broad H$\beta$ 
($i=1,~\dots,~4$), we have $f_{b,~i}~\ge~0$, $\sigma_{b,~i}~\ge~400{\rm km/s}$ and 
$4700\textsc{\AA}~\le~\lambda_{0b,~i}~\le~4980\textsc{\AA}$, and have zero as staring values of 
$f_{b,~i}$, have 1000${\rm km/s}$ as starting values of $\sigma_{b,~i}$, and have 4750\AA, 4800\AA, 
4890\AA, 4930\AA~ as starting values of $\lambda_{0b,~i}$. For the parameters of the extended 
components of narrow H$\beta$ and [O~{\sc iii}] doublet ($k=1,~\dots,~3$), we have $f_{e,~k}~\ge~0$, 
$\sigma_{e,~k}$ larger than those of corresponding narrow components, 
$\lambda_{0T,~k}-30\textsc{\AA}~\le~\lambda_{0e,~k}~\le~\lambda_{0T,~k}+30\textsc{\AA}$, and have 
zero as starting values of $f_{e,~k}$, have 600${\rm km/s}$ as starting values of $\sigma_{e,~k}$, 
and have $\lambda_{0T,~k}$ as the starting values of $\lambda_{0e,~k}$. For the power law component, 
we have $\alpha_2~\ge~0$ and $-5~\le~\beta_2~\le~1$. Meanwhile, the flux ratio of [O~{\sc iii}] 
doublet is fixed to the theoretical values of 3, and the components from the [O~{\sc iii}] doublet 
have the same redshift and the same second moment.

	Then, through the Levenberg-Marquardt least-squares minimization method, the best-fitting 
results and corresponding residuals (line spectrum minus the best fitting results) to 
the emission lines can be well determined by the model functions above and shown in Figure~\ref{line}. 
The values of $\chi^2$ (summed squared residuals divided by degree of freedom) for the best fitting 
results to the emissions lines by the model functions above can be well determined and listed in 
Table~1.As discussions above on criteria to collect double-peaked BLAGN from the sample 
of \citet{liu19}, the candidate for double-peaked BLAGN in \citet{liu19} are not considered, if there 
are no narrow H$\beta$ or no apparent broad H$\beta$ in SDSS spectra. Fig.~\ref{l2ine} shows one 
example on the shown spectrum without apparent narrow H$\beta$ in SDSS 1347-52823-0296, and one 
example on the shown spectrum without apparent broad H$\beta$ in SDSS 1867-53317-0344, after 
subtractions of host galaxy contributions.

       Based on the shown results in Figure~\ref{line}, three points are noted. 
First and foremost, after the emission lines are measured, broad Balmer emission lines have slightly 
different line profiles, such as the results shown in 0725-52258-0510 with broad H$\alpha$ described 
by four broad Gaussian components but with broad H$\beta$ described by only two broad Gaussian 
components. The difference could be due to worse mixed broad H$\beta$ emissions with extended 
[O~{\sc iii}] emissions. Therefore, in the manuscript, properties of the disk-like BLRs are mainly 
determined through the double-peaked broad H$\alpha$. Besides, as the shown residuals in 
Figure~\ref{spec} and the best-fitting results and the corresponding residuals to emission lines 
around H$\beta$ in Figure~\ref{line}, it is not necessary to consider the optical Fe~{\sc ii} emission 
lines in the 35 double-peaked BLAGN, besides the three double-peaked BLAGN SDSS 0966-52642-0499, SDSS 
2123-53793-0443 and SDSS 2783-54524-0517. Similar as what we have done in \citet{zh21}, model functions 
including optical Fe~{\sc ii} templates \citep{kp10} and probable He~{\sc ii} component have been 
considered to describe the emissions within rest wavelength from 4400 to 5600\AA~ in the line spectra of 
SDSS 0966-52642-0499, SDSS 2123-53793-0443 and SDSS 2783-54524-0517, of which determined optical 
Fe~{\sc ii} emissions shown as dashed cyan lines in Figure~\ref{line}. Last but not the least, there 
are extended components in narrow Balmer emission lines in the model functions, however, 
the extended narrow Balmer components are only detected in SDSS 0721-52228-0600. 
The core and the extended components in narrow Balmer lines have line widths (second moment) 
about ${\rm 98km/s}$ and ${\rm 230km/s}$ in SDSS 0721-52228-0600. Due to the line width (second moment) 
about ${\rm 240km/s}$ of the extended [O~{\sc iii}] component larger than the line width (second moment) 
about ${\rm 230km/s}$ of the extended components in narrow Balmer lines in SDSS 0721-52228-0600, therefore, 
the determined extended components in narrow Balmer lines are accepted from narrow Balmer emission regions in 
SDSS 0721-52228-0600. Although the multiple Gaussian components can not provide physical properties of the 
expected disk-like BLRs, the results in Figure~\ref{line} can provide more accurate measurements 
of the narrow emission lines, especially the line luminosity of [O~{\sc iii}]$\lambda5007$\AA. The 
determined narrow emission lines around H$\alpha$ can be subtracted from the line spectrum to obtain 
the pure double-peaked broad H$\alpha$ which will be described by the elliptical accretion disk model.

       Finally, through the Levenberg-Marquardt least-squares minimization method, the pure 
double-peaked broad H$\alpha$ are best fitted by the elliptical accretion disk model 
and shown in Figure~\ref{disk}. When the elliptical accretion disk model is applied, the seven model 
parameters have restrictions as follows. The inner inner boundary $r_0$ is larger than 15${\rm R_G}$ 
and smaller than 1000${\rm R_G}$. The out boundary $r_1$ is larger than $r_0$ and smaller than 
$10^6{\rm R_G}$. The inclination angle $i$ of disk-like BLRs has $\sin(i)$ larger than 0.05 and 
smaller than 0.95. The eccentricity $e$ is larger than 0 and smaller than 1. The orientation angle 
$\phi_0$ of elliptical rings is larger than 0 and smaller than $2\pi$. The local broadening velocity 
$\sigma_L$ is larger than 10${\rm km/s}$ and smaller than $10^4{\rm km/s}$. Here, two 
points are noted on properties of the local broadening velocity. On the one hand, it is not totally 
clear on the origin of the local broadening velocity. As discussed in \citet{ch89}, the most likely 
interpretation to the local broadening is due to the effects of electron scattering in a photoionized 
atmosphere of the disk. On the other hand, as the reported model fitting results to the double-peaked 
broad emission lines, such as in \citet{ch89, el95, st03} etc., the local broadening velocity could 
be round a few hundreds to a few thousands of kilometers per second. Therefore, in the manuscript, 
the upper limit $10^4{\rm km/s}$ is accepted to the local broadening velocity in the model. The 
line emissivity slope $q$ ($f_r~\propto~r^{-q}$) is larger than -7 and smaller than 7. Fortunately, 
the double-peaked broad H$\alpha$ of the 38 double-peaked BLAGN can be well described 
by the elliptical accretion disk model, therefore, the other accretion disk models with improved 
subtle structures are not discussed any more in the manuscript. The determined seven model parameters 
and corresponding uncertainties are listed in Table~2 for the double-peaked broad H$\alpha$ in 
each double-peaked BLAGN. Here, the uncertainty of each model parameter is the formal 
1sigma error computed from the covariance matrix, through the Levenberg-Marquardt least-squares 
minimization technique (the MPFIT package\footnote{\url{https://pages.physics.wisc.edu/~craigm/idl/cmpfit.html}}) 
\citep{mc09}. The determined $\chi^2$ values for the best descriptions 
to the double-peaked broad H$\alpha$ by the elliptical accretion disk model are listed in Table~1.

       Based on the procedures above, the measured continuum intensity at 5100\AA, the measured 
[O~{\sc iii}]$\lambda5007$\AA~ line flux, the measured fluxes of narrow Balmer emission lines and the 
determined inclination angle of the disk-like BLRs are listed in Table~1. Certainly, based on each 
measured inclination angle of the disk-like BLRs in central accretion disk and the redshift, the 
expected upper limit $R_{NLRs,u}$ of NLRs sizes can be calculated and also listed in the final 
column of Table~1.


	Before proceeding further, two points are noted. On the one hand, there are five 
nearby double-peaked BLAGN reported and discussed in \citet{sb17}. Among the five double-peaked 
BLAGN, NGC5548 has been observed in SDSS with PLATE-MJD-FIBERID=2127-53859-0085. However, there 
are no apparent bumps in the line profile of broad H$\alpha$. Therefore, NGC 5548 is not included in the 
sample of candidates for double-peaked BLAGN in \citet{liu19} nor included in our final sample of 
low redshift double-peaked BLAGN. However, based on the reported inclination angle 19\degr~ 
($\sin(i)\sim0.3256$) of disk emission regions in NGC 5548\ in \citet{sb17}, upper limit of NLRs 
size of NGC 5548 can be estimated as $R_{NLRs,u}~\sim~1550pc$. Based on the shown SDSS spectra and 
the best fitting results to the emission lines of NGC5548 in Figure~\ref{n5548}, [O~{\sc iii}] line 
flux $f_{[O~\textsc{iii}]}\sim44813\times10^{-17}{\rm erg/s/cm^2}$ is accepted in NGC5548. In one 
word, NGC5548 is not included in our final sample, however, the estimated $R_{NLRs,u}$ and 
$f_{[O~\textsc{iii}]}$ (corresponding \o3~ line luminosity $2.67\times10^{41}{\rm erg/s}$) will 
be applied in the following results. On the other hand, when the elliptical accretion disk model 
is applied, the model parameter $r_1$ and $\sin(i)$ have probably similar effects on features around 
the peaks of double-peaked line profiles, therefore it is necessary to check whether the parameter 
$r_1$ and the parameter $\sin(i)$ can be clearly and solely determined through the model applied 
to describe the double-peaked broad lines. In other words, it is necessary to determine that the 
parameters $\sin(i)$ and the other model parameters are not completely degenerate together. Here, 
a simple procedure is applied as follows. Based on randomly collected values of the model parameters 
within their limits as described above (input values of the model parameters) in the elliptical 
accretion disk model, 400 simulating double-peaked broad H$\alpha$ are created with considerations 
of signal-to-noise about 15. Left panel of Figure~\ref{tsin} shows an example on the simulating 
double-peaked broad H$\alpha$. Then, the same procedure considering the elliptical accretion disk 
model is applied to fit the simulating double-peaked broad H$\alpha$ through the Levenberg-Marquardt 
least-squares minimization method with [200$R_G$, 1000$R_G$, 0.5, 2, 1000${\rm km/s}$, 0.2, 0.] as 
the starting values of the model parameters, leading to the measured values of the model parameters. 
An example on the best-fitting results to the simulating double-peaked broad H$\alpha$ is shown in 
the left panel of Figure~\ref{tsin}. If the model parameter of $\sin(i)$ and the other model 
parameters are degenerate together, the input values of $\sin(i)$ should be quite different from 
the measured values of $\sin(i)$. However, as shown in the top second panel of Figure~\ref{tsin}, there 
is a strong linear correlation with Spearman Rank correlation coefficient 0.928 ($P_{null}<10^{-15}$) 
between the input $\sin(i)$ and the measured $\sin(i)$. Moreover, besides the model 
parameter of $\sin(i)$, there are apparent linear correlations between the input model parameters 
and the re-measured model parameters determined by best fitting results to the simulating double-peaked 
broad H$\alpha$, with Spearman Rank correlation coefficients about 0.66 with $P_{null}<10^{-15}$, 
about 0.46 with $P_{null}<10^{-15}$, and 0.83 with $P_{null}<10^{-15}$, about 0.88 with $P_{null}<10^{-15}$, 
and 0.71 with with $P_{null}<10^{-15}$ for the correlations between between input $r_0$ and re-measured $r_0$, 
between input $r_1$ and re-measured $r_1$, between input $e$ and re-measured $e$, between input $q$ and 
re-measured $q$, between input $\sigma_L$ and re-measured $\sigma_L$, respectively. Therefore, not only 
the model parameter $\sin(i)$ but also the other model parameters can be solely determined in the elliptical 
accretion disk model, as long as the double-peaked line profiles are clean enough. 


\begin{figure*}[htp]
\centering\includegraphics[width = 18cm,height=20cm]{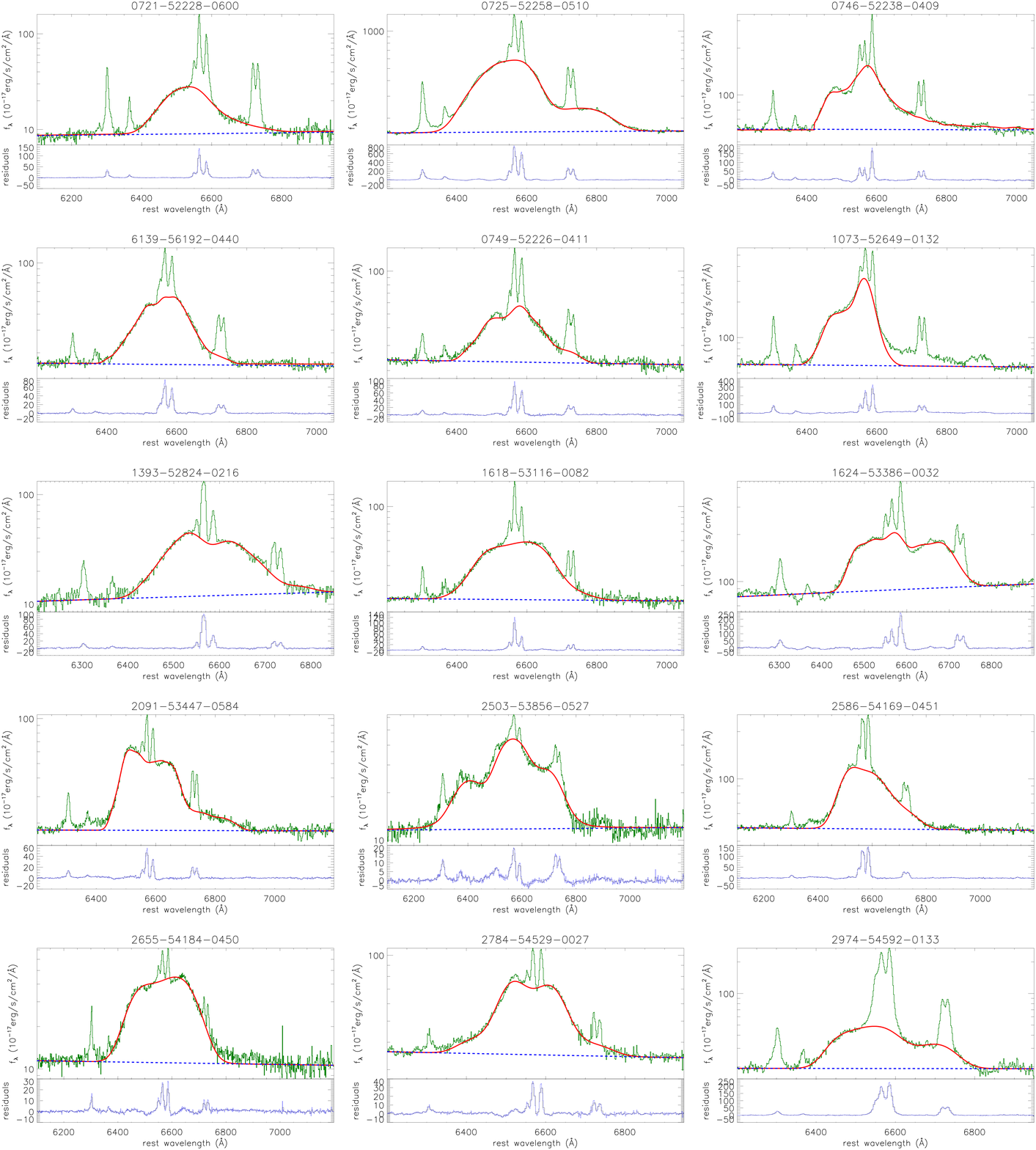}
\caption{Elliptical accretion disk model determined the best descriptions to the double-peaked 
broad H$\alpha$. In each panel, solid line in dark green shows the line spectrum after subtractions 
of the contributions of stellar lights, solid red line shows the elliptical accretion disk model 
determined best descriptions to the double-peaked broad H$\alpha$, dashed blue line shows the power 
law continuum emissions similar as the one shown in Figure~\ref{line}. Bottom region 
of each panel shows the residuals calculated by line spectrum minus sum of the best descriptions 
to the double-peaked broad H$\alpha$ plus the power law continuum emissions. {\bf In order to 
show clearly spectral broad emission line features, the Y-axis is in logarithmic coordinate for the 
plots in top region of each panel.}
}
\label{disk}
\end{figure*}

\setcounter{figure}{3}

\begin{figure*}[htp]
\centering\includegraphics[width = 18cm,height=22cm]{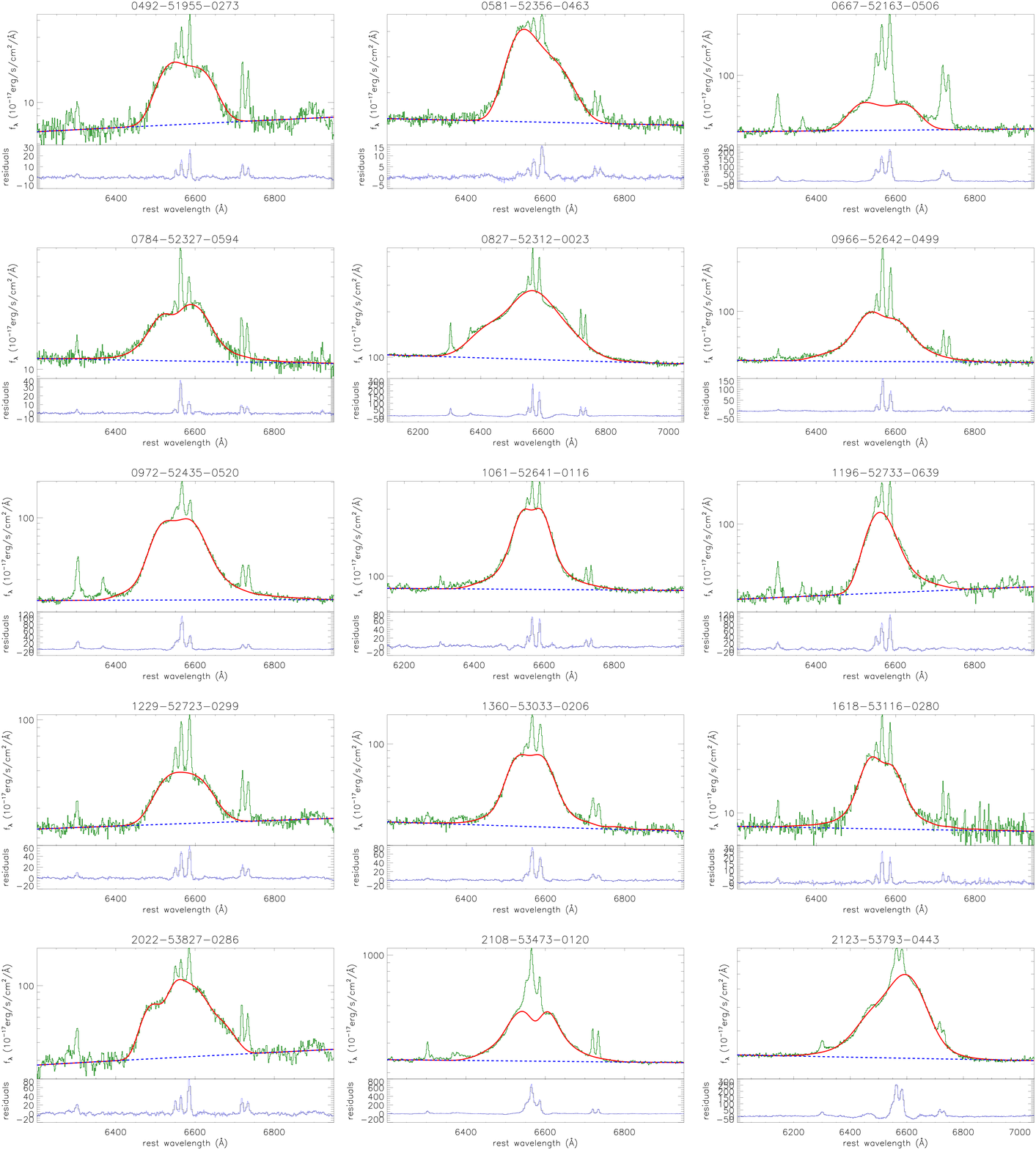}
\caption{--continued.}
\end{figure*}

\setcounter{figure}{3}

\begin{figure*}[htp]
\centering\includegraphics[width = 18cm,height=11cm]{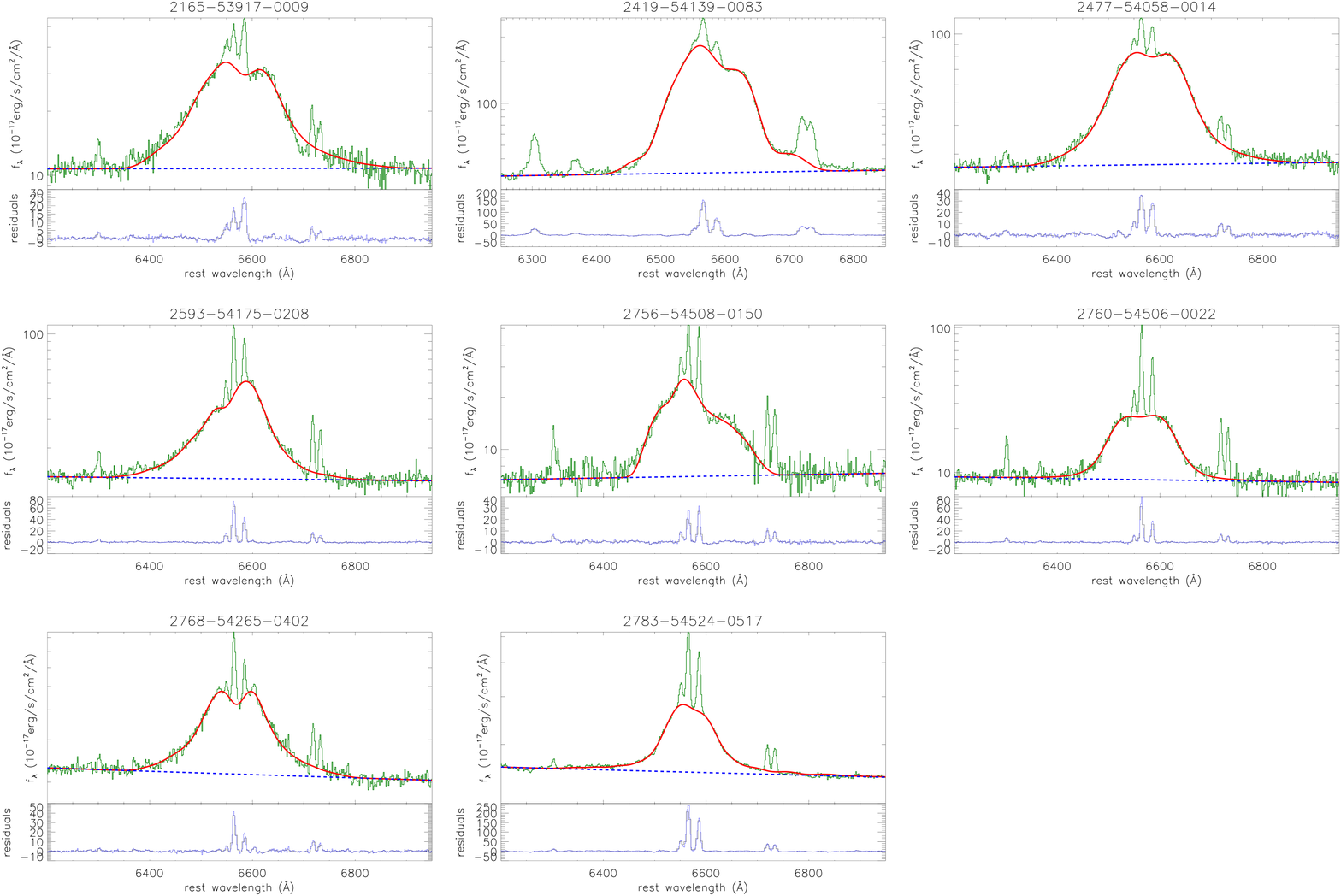}
\caption{--continued.}
\end{figure*}

\begin{figure*}[htp]
\centering\includegraphics[width = 18cm,height=4cm]{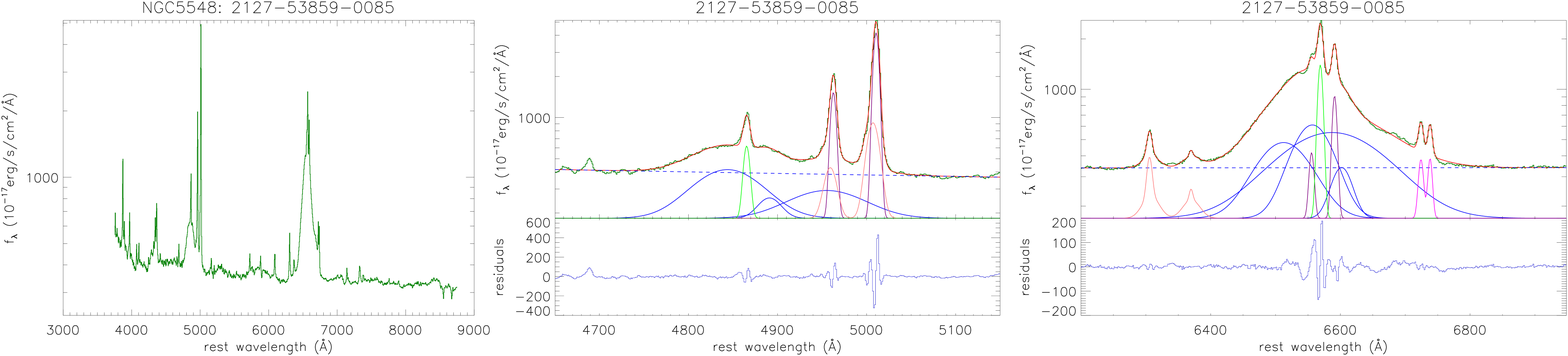}
\caption{SDSS spectrum of NGC5548 (left panel) and the best fitting results to the 
emission lines around H$\beta$ (middle panel) and around H$\alpha$ (right panel). Symbols and 
line styles in the middle and the right panel have the same meanings as those in Figure~\ref{line}.}
\label{n5548}
\end{figure*}

\begin{figure*}[htp]
\centering\includegraphics[width = 18cm,height=6cm]{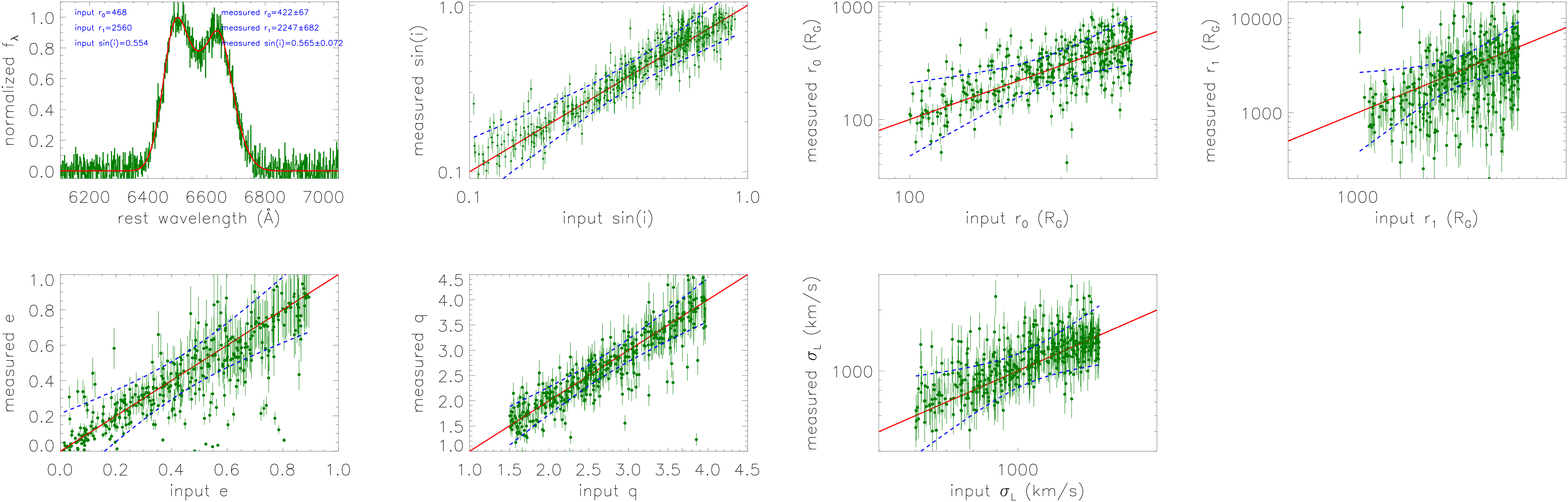}
\caption{Top left panel shows an example on the simulating double-peaked broad H$\alpha$ 
(dark green) and the best-fitting results (red). In left panel, the input and measured values of $r_0$, 
$r_1$ and $\sin(i)$ are marked in blue characters. The other panels show the correlations 
between the six input model parameters and the six re-measured model parameters with corresponding 
uncertainties determined by the best-fitting results to the simulating double-leaked broad H$\alpha$. In 
the last six panels, solid red lines and dashed blue lines show the X=Y and the corresponding 99.9999\% 
confidence bands to X=Y.
}
\label{tsin}
\end{figure*}

\begin{figure*}[htp]
\centering\includegraphics[width = 15cm,height=10cm]{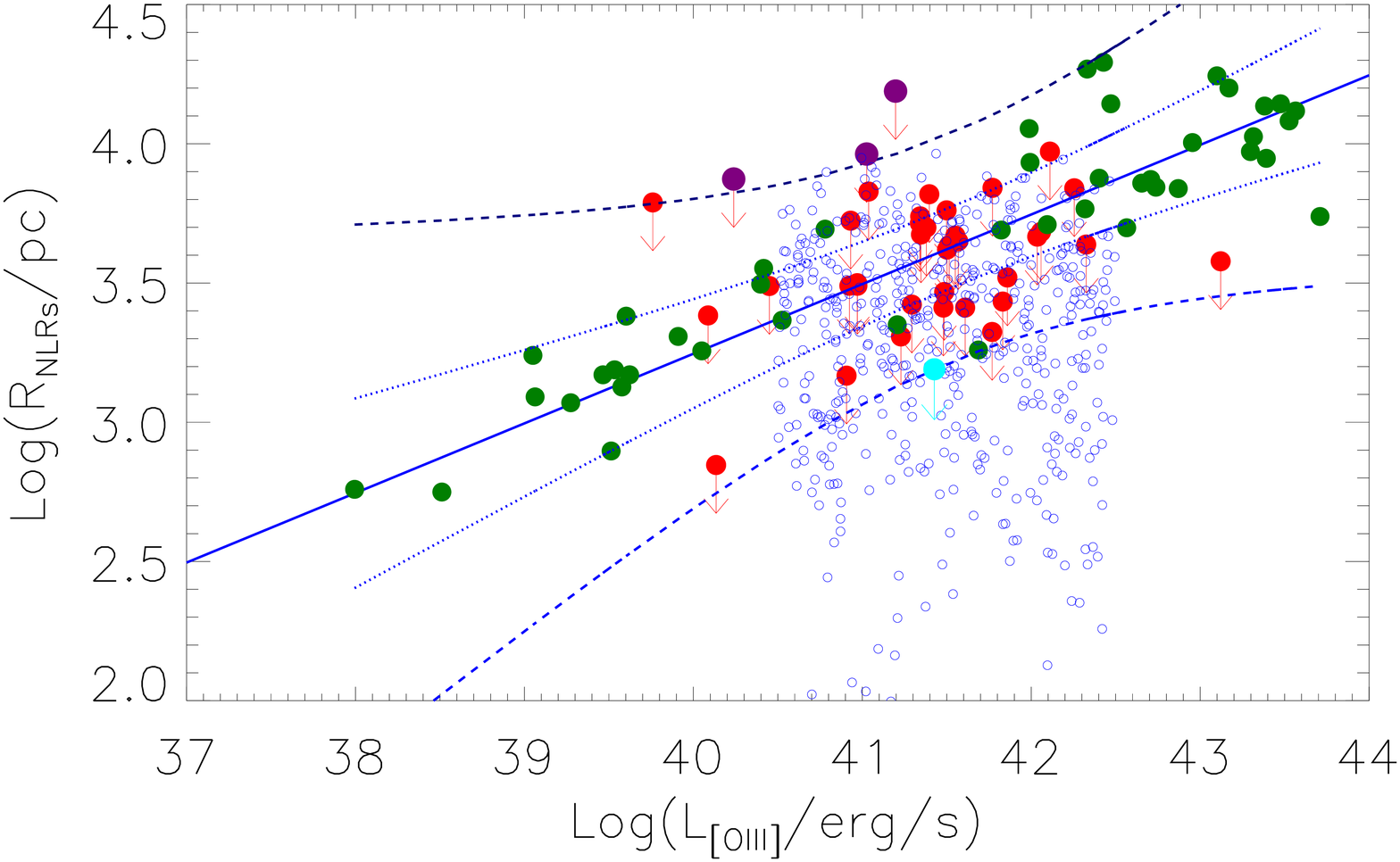}
\caption{Properties of $R_{NLRs,u}$ of the 38 double-peaked BLAGN (shown as solid red 
circles) in the plane of NLRs size versus [O~{\sc iii}] luminosity. Solid circles in dark green are 
the results for type-2 AGN with measured NLRs sizes discussed in \citet{liu13}, solid blue line shows 
the empirical R-L relation reported in \citet{liu13}. Dotted and dashed blue lines show the corresponding 
90\% and 99.9999\% confidence bands of the empirical R-L relation. The three outliers, SDSS 
0581-52356-0463, SDSS 1196-52733-0639 and SDSS 2756-54508-0150, are marked as solid purple circles in 
the figure. Solid cyan circle represents the results for NGC5548. The 500 blue open circles 
represent the simulating results based on the created $\log(L_{O~{\textsc{iii}}})$ line luminosity and 
the randomly created $\sin(i)$ and redshift.}
\label{res}
\end{figure*}

\begin{table*}
\caption{Parameters of the 38 double-peaked BLAGN}
\begin{center}
\begin{tabular}{ccccccccccccc}
\hline\hline
PMF  & $z$  &  $f_{[O~{\textsc{iii}}]}$  & $f_{H\alpha}$  &  $f_{H\beta}$ & $E(B-V)$ & $\sin(i)$ & 
	$\chi^2_{H\beta}$ & $\chi^2_{H\alpha}$ & $\chi^2_{disk}$ & $f_{5100}$  & $R$ \\
\hline
0721-52228-0600 & 0.058 &  2699$\pm$56    & 886$\pm$22 &  307$\pm$37 & \dots & 0.325$\pm$0.013  & 1.14  
	& 1.78  & 1.67  & 9.34 & 5811 \\
0725-52258-0510 & 0.047 &  25564$\pm$350  & 8601$\pm$148 & 2342$\pm$39 & 0.145 &  0.318$\pm$0.002  & 5.16 
	&  1.04 &  0.51 & 168.83 & 4766 \\
0746-52238-0409 & 0.046 &  2656$\pm$66 & 442$\pm$25  &  99$\pm$10 & 0.31 & 0.289$\pm$0.001 & 1.06 
	&  0.96 &  1.65  & 55.59 & 5118\\
6139-56192-0440* &  0.096 &  1946$\pm$73 & 823$\pm$23 & 189$\pm$8 & 0.287 & 0.555$\pm$0.028  & 1.18 
	&  0.93  & 1.02 & 17.85 & 3868 \\
\hline\hline
0749-52226-0411 & 0.096 & 2310$\pm$54 & 909$\pm$37 & 228$\pm$9 & 0.216 & 0.417$\pm$0.012 & 2.23 
	& 1.43 & 0.22  & 25.67 & 7685 \\  
1073-52649-0132 &  0.026 & 10887$\pm$226 &   2262$\pm$127 &  745$\pm$39 & \dots & 0.387$\pm$0.021 & 1.34 
	& 0.49 & 0.06  & 70.154 & 2134 \\
1393-52824-0216 &  0.041 &  1415$\pm$60  & 1024$\pm$26 &   285$\pm$71 & 0.126 & 0.394$\pm$0.035 &  1.08 
	& 0.92 & 0.11 &  8.561 & 3342\\
1618-53116-0082 &  0.073  & 2778$\pm$56  &  675$\pm$20  &  251$\pm$9  & \dots & 0.471$\pm$0.055  & 1.37 
	& 0.65 & 0.88 &  18.627 & 5093 \\
1624-53386-0032 & 0.0075  & 2265$\pm$133  & 1126$\pm$78  &  206$\pm$21  & 0.488 & 0.332$\pm$0.009 & 0.97 
	& 1.09 & 1.10 &  56.58 & 707\\
2091-53447-0584 &  0.073  & 770$\pm$69  &  480$\pm$18  &  151$\pm$28 & 0.02 & 0.312$\pm$0.002 & 1.01 
	& 0.81 & 1.42 &  20.31 & 7688 \\
2503-53856-0527 &  0.071 &  465$\pm$71   & 235$\pm$16  & 64$\pm$8  & 0.147 & 0.663$\pm$0.059 & 0.96 
	& 0.76 & 1.03  & 12.88 & 3514\\
2586-54169-0451 &  0.084  & 2175$\pm$131 &  1955$\pm$90  &  406$\pm$23 & 0.378 & 0.254$\pm$0.005 & 0.73 
	& 0.83 & 1.39  & 54.78 & 10950\\
2655-54184-0450 &  0.084  & 596$\pm$63  &  227$\pm$18 &   49$\pm$8  & 0.342 & 0.412$\pm$2.671$^{**}$  &  0.86 
	& 0.79 & 1.25  & 12.55 & 6751\\
2784-54529-0027 &  0.083  & 916$\pm$9   &  355$\pm$15  &  91$\pm$8  & 0.202 & 0.805$\pm$0.125 & 1.23 
	& 1.22 & 1.82  & 44.47 & 3411\\
2974-54592-0133 &  0.043  & 7513$\pm$216  & 1562$\pm$214 &  687$\pm$25 & \dots & 0.294$\pm$0.022  & 2.25 
	& 0.61 & 0.58  & 22.77 & 4704 \\
\hline\hline
0492-51955-0273 & 0.039 & 162$\pm$178 & 80$\pm$9 & 41$\pm$9 & \dots & 0.188$\pm$1.185$^{**}$ & 0.82 & 
	0.86 & 1.01 & 4.14 & 6653 \\
0581-52356-0463 & 0.091 & 217$\pm$24 & 88$\pm$11 & 18$\pm$5 & 0.391 & 0.165$\pm$0.291$^{**}$ & 0.97 & 
	0.85 & 1.19 & 16.73 & 18350 \\
0667-52163-0506 & 0.042 & 1952$\pm$53 & 957$\pm$33 & 302$\pm$18 & 0.018 & 0.235$\pm$1.754$^{**}$ & 0.82 & 
	0.91 & 1.12 & 38.56 & 5744 \\
0784-52327-0594 &  0.098 &  331$\pm$16 & 269$\pm$7 & 60$\pm$8  &  0.318 & 0.575$\pm$0.041 & 0.97 & 0.92  & 
	 1.03  &  13.28  & 5697 \\
0827-52312-0023 & 0.092 & 6151$\pm$92 & 1383$\pm$28 & 396$\pm$16 & 0.1 & 0.371$\pm$0.017 & 1.75 & 0.91 &
	 1.86 &  140.58 & 8256 \\
0966-52642-0499 & 0.096 & 946$\pm$39 & 1099$\pm$21 & 230$\pm$10 & 0.372 & 0.806$\pm$0.183 & 0.98 & 0.68 &
	 0.73 & 57.88 & 3431 \\
0972-52435-0520 & 0.094 & 2313$\pm$39 & 685$\pm$31 & 210$\pm$11 &
	 0.043 & 0.376$\pm$0.011 & 2.11 & 0.84 & 1.15 & 20.64 & 8334 \\
1061-52641-0116 & 0.041 & 1141$\pm$52 & 553$\pm$39 & 80$\pm$15 &
	 0.69 & 0.471$\pm$0.023 & 1.28 & 0.63 & 0.82 & 95.66 & 2796 \\
1196-52733-0639 & 0.022 & 1289$\pm$66 & 576$\pm$41 & 172$\pm$36 &
	 0.063 & 0.089$\pm$0.335$^{**}$ & 0.72 & 0.46 & 0.68 & 26.21 & 7828 \\
1229-52723-0299 & 0.034 & 323$\pm$38 & 324$\pm$18 & 68$\pm$10 &
	 0.37 & 0.329$\pm$0.018 & 0.89 & 0.67 & 0.73 & 18.49 & 3302 \\
1360-53033-0206 & 0.067 & 1877$\pm$67 & 700$\pm$36 & 192$\pm$19 &
	 0.138 & 0.461$\pm$0.044 & 1.61 & 0.64 & 0.59 & 36.75 & 4755 \\
1618-53116-0280 & 0.074 & 388$\pm$19 & 150$\pm$7 & 39$\pm$5 &
	 0.183 & 0.667$\pm$0.215 & 0.86 & 0.92 & 0.89 & 9.29 & 3648 \\
2022-53827-0286 & 0.023 & 571$\pm$126 & 240$\pm$35 & 63$\pm$15 &
	 0.18 & 0.287$\pm$0.012 & 0.59 & 0.35 & 0.33 & 20.58 & 2540 \\
2108-53473-0120 & 0.023 & 6708$\pm$162 & 2310$\pm$162 & 800$\pm$72 &
	 \dots & 0.473$\pm$0.014 & 1.59 & 0.41 & 0.62 & 249.40 & 1541 \\
2123-53793-0443 & 0.077 & 3786$\pm$227 & 3531$\pm$252 & 360$\pm$92 &
	 0.994 & 0.577$\pm$0.095 & 0.71 & 0.58 & 1.02 & 249.45 & 4397 \\
2165-53917-0009 & 0.085 & 391$\pm$37 &  170$\pm$23&  35$\pm$11 &
	 0.384 & 0.478$\pm$0.024 & 0.94 & 0.83 & 0.97 & 10.37 & 5892 \\
2419-54139-0083 & 0.047 & 4297$\pm$118 & 1677$\pm$115 & 380$\pm$62 &
	 0.302 & 0.655$\pm$0.104 & 1.52 & 0.41 & 0.51 & 25.21 & 2314 \\
2477-54058-0014 & 0.047 & 587$\pm$191 & 408$\pm$39 & 67$\pm$12 &
	 0.582 & 0.522$\pm$0.037 & 0.81 & 0.62 & 0.64 & 25.24 & 2904 \\
2593-54175-0208 & 0.099 & 440$\pm$27 & 477$\pm$12 & 114$\pm$7 &
	0.258 & 0.416$\pm$0.018 & 0.82 & 0.64 & 0.64 & 15.28 & 7960 \\
2756-54508-0150 & 0.082 & 637$\pm$19 & 162$\pm$7 & 53$\pm$8 &
	 \dots & 0.252$\pm$0.009 & 0.93 & 0.86 & 0.96 & 6.22 & 10759 \\
2760-54506-0022 & 0.078 & 1429$\pm$26 & 439$\pm$10 & 125$\pm$7 &
	 0.108 & 0.857$\pm$0.767 & 1.43 & 0.90 & 0.86 & 10.21 & 3001 \\
2768-54265-0402 & 0.091 & 727$\pm$29 & 278$\pm$9 & 78$\pm$7 &
	 0.118 & 0.463$\pm$0.024 & 0.93 & 0.91 & 0.97 & 27.75 & 6539 \\
2783-54524-0517 & 0.055 & 2197$\pm$82 & 1932$\pm$47 & 368$\pm$24 &
	 0.455 & 0.591$\pm$0.034 & 1.87 & 0.62 & 0.58& 192.70 & 3019 \\
\hline
\end{tabular}\\
\end{center}
Notice: The first and second column show the information of PLATE-MJD-FIBERID and redshift, the third to 
fifth columns show the line fluxes of [O~{\sc iii}]$\lambda5007$\AA, narrow H$\alpha$ and narrow H$\beta$, 
in units of $10^{-17}{\rm erg/s/cm^2}$, the sixth column shows the corresponding E(B-V) calculated from the 
flux ratio of narrow H$\alpha$ to narrow H$\beta$, the seventh column shows elliptical accretion disk model 
determined $\sin(i)$, the eighth, ninth and tenth column show the values of $\chi^2$ for the best 
descriptions to the emission lines around H$\beta$ and around H$\alpha$ by multiple Gaussian functions, 
and to the broad double-peaked H$\alpha$ by the elliptical accretion disk model, the eleventh column shows 
the continuum intensity at 5100\AA~ in the units of $10^{-17}{\rm erg/s/cm^2/\textsc{\AA}}$, 
and the final column shows the estimated upper limit of NLRs sizes $R_{NLRs,u}$ in the units of $pc$. \\
The first four rows are for the double-peaked BLAGN from the sample of \citet{st03}. The fifth 
row to the fifteenth row are for the double-peaked BLAGN collected from SDSS quasars. The other rows are for 
the double-peaked BLAGN collected from the sample of \citet{liu19}\\
SDSS 6139-56192-0440 has its SDSS spectrum from the eBOSS with fiber diameter of 2 arcseconds, and the 
value $R_{NLRs,u}$ of SDSS 6139-56192-0440 is estimated by the fiber radius of 1 arcseconds.\\
$^{**}$ means that the value is the one determined by the Levenberg-Marquardt least-squares minimization 
method, although the uncertainty is larger than the value.
\end{table*}

\begin{table*}
\caption{Parameters of the disk-like BLRs of the 38 double-peaked BLAGN}
\begin{tabular}{cccccccc}
\hline\hline
PMF  & $r_0$  & $r_1$  &  $\sin(i)$ & $q$ & $\sigma_L$ & $e$ & $\phi_0$  \\
     & $R_G$  & $R_G$  &            &     &  $km/s$    &     &   \\
\hline
0721-52228-0600 & 94$\pm$12 & 3450$\pm$560 & 0.325$\pm$0.013 & 1.98$\pm$0.07 & 1170$\pm$110 & 0.20$\pm$0.02 & 0.32$\pm$0.03 \\
0725-52258-0510 & 61$\pm$2 & 1330$\pm$640 &  0.318$\pm$0.002 & 4.11$\pm$0.15 & 1740$\pm$40 & 0.31$\pm$0.01 & 1.81$\pm$0.03 \\
0746-52238-0409 & 29$\pm$1 & 4826$\pm$420 & 0.289$\pm$0.001 & 2.23$\pm$0.01 & 650$\pm$35 & 0.78$\pm$0.01 & 2.47$\pm$0.02 \\
6139-56192-0440 & 476$\pm$55 & 5175$\pm$670 & 0.555$\pm$0.028 & 2.03$\pm$0.04 & 460$\pm$50 & 0.70$\pm$0.01 & 2.91$\pm$0.01 \\
\hline\hline
0749-52226-0411 & 153$\pm$14 & 53329$\pm$25296 & 0.417$\pm$0.012 & 3.87$\pm$0.07 & 875$\pm$60 & 0.58$\pm$0.12 & 0.24$\pm$0.03 \\ 
1073-52649-0132 & 235$\pm$77 & 3773$\pm$14627 & 0.387$\pm$0.021 & 3.05$\pm$0.44 & 770$\pm$146 & 0.67$\pm$0.22 & 0.58$\pm$0.04 \\
1393-52824-0216 & 159$\pm$27 & 1911$\pm$1354 & 0.394$\pm$0.035 & 3.61$\pm$1.38 & 1140$\pm$180 & 0.26$\pm$0.06 & 0.01$\pm$0.02 \\
1618-53116-0082 & 175$\pm$46 & 3485$\pm$2526 & 0.471$\pm$0.055 & 3.05$\pm$0.39 & 1520$\pm$260 & 0.24$\pm$0.06 & 0.42$\pm$0.11 \\ 
1624-53386-0032 & 148$\pm$8 & 4633$\pm$678  & 0.332$\pm$0.009 & 4.46$\pm$0.16 & 900$\pm$20 & 0.57$\pm$0.03 & 0.01$\pm$0.01 \\ 
2091-53447-0584 & 94$\pm$2 & 2749$\pm$110 &  0.312$\pm$0.002 & 2.78$\pm$0.04 & 770$\pm$30 & 0.01$\pm$0.01 & 0.23$\pm$0.02 \\
2503-53856-0527 & 51$\pm$17 & 15063$\pm$3562 &  0.663$\pm$0.059 & 3.47$\pm$0.21 & 1610$\pm$45 & 0.27$\pm$0.03 & 2.04$\pm$0.04 \\
2586-54169-0451 & 64$\pm$36 & 9879$^{**}$ & 0.254$\pm$0.005 & 4.76$\pm$4.96 & 1620$\pm$460 & 0.42$\pm$0.24 & 0.74$\pm$0.98 \\
2655-54184-0450 & 267$^{**}$ &  2343$^{**}$ & 0.412$\pm$2.671$^{**}$ & 4.07$^{**}$ & 1980$^{**}$ & 0.28$^{**}$ & 0.49$^{**}$ \\
2784-54529-0027 & 888$\pm$316 & 14900$^{**}$ &  0.805$\pm$0.125 & 3.41$\pm$1.08 & 1300$\pm$310 & 0.73$\pm$0.61 & 3.13$\pm$0.02 \\
2974-54592-0133 & 76$\pm$66 & 4101$^{**}$ & 0.294$\pm$0.022 & 4.32$\pm$0.42 & 1530$\pm$115 & 0.36$\pm$0.36 & 1.39$\pm$0.38 \\
\hline\hline
0492-51955-0273 & 383$^{**}$   & 540$^{**}$ &  0.188$^{**}$ 
	&  4.47$^{**}$ &  1395$\pm$450 &  0.26$^{**}$ &  2.48$^{**}$ \\
0581-52356-0463 & 154$\pm$104  & 729$^{**}$  &  0.165$^{**}$
	& 4.24$^{**}$  & 1515$\pm$40  & 0.56$^{**}$ &  2.01$^{**}$ \\
0667-52163-0506 & 177$^{**}$   & 317$^{**}$  & 0.235$^{**}$ 
	& 3.85$^{**}$  & 1410$\pm$372  & 0.29$^{**}$  & 1.63$^{**}$  \\
0784-52327-0594 & 78$\pm$27   & 4548$\pm$622  & 0.575$\pm$0.041 
        & 1.13$\pm$0.08 & 934$\pm$62  & 0.19$\pm$0.01  & 0.21$\pm$0.01  \\
0827-52312-0023 & 91$\pm$100  & 310$\pm$277  & 0.371$\pm$0.017 
	& 2.32$^{**}$  & 2451$\pm$92  & 0.95$\pm$0.01  & 1.68$\pm$0.08  \\
0966-52642-0499 & 581$\pm$287   & 22827$\pm$10699  & 0.806$\pm$0.183
	& 1.59$\pm$0.07  & 1174$\pm$44  & 0.001$^{**}$  & 2.41$\pm$0.11  \\
0972-52435-0520 & 87$\pm$9   & 5396$\pm$426  & 0.376$\pm$0.011 
	& 1.33$\pm$0.02  & 1166$\pm$30  & 0.04$\pm$0.01  & 1.98$\pm$0.07  \\
1061-52641-0116 & 164$\pm$23   & 9663$\pm$1110  & 0.471$\pm$0.023 
	& 1.45$\pm$0.02  & 930$\pm$34  &  0.18$\pm$0.02 & 2.32$\pm$0.06  \\
1196-52733-0639 & 77$\pm$202   & 10914$^{**}$  & 0.089$^{**}$ 
	& 1.83$\pm$2.10  & 1367$\pm$193  & 0.00$^{**}$  & 2.09$^{**}$  \\
1229-52723-0299 & 749$\pm$759   & 3390$\pm$4338  & 0.329$\pm$0.018 
	& 2.13$^{**}$  & 1450$\pm$94  & 0.00$^{**}$  & 2.36$^{**}$  \\
1360-53033-0206 & 54$\pm$11   & 8729$\pm$2078  & 0.461$\pm$0.044 
	& 1.16$\pm$0.05  & 986$\pm$44  & 0.03$\pm$0.02  & 2.18$\pm$0.04  \\
1618-53116-0280 & 172$\pm$115   & 23035$\pm$15353  & 0.667$\pm$0.215
	& 1.28$\pm$0.07  & 895$\pm$94  & 0.00$^{**}$  & 2.38$\pm$0.24  \\
2022-53827-0286 & 196$\pm$31   & 4949$\pm$1532  & 0.287$\pm$0.012 
	&  2.70$\pm$0.08 & 817$\pm$49  & 0.57$\pm$0.08  & 2.16$\pm$0.04  \\
2108-53473-0120 & 241$\pm$19   & 5795$\pm$374  & 0.473$\pm$0.014 
	& 1.46$\pm$0.03  & 800$\pm$25  & 0.09$\pm$0.01  & 2.25$\pm$0.01  \\
2123-53793-0443 & 103$\pm$37   & 4094$\pm$1532  & 0.577$\pm$0.096 
	& 2.16$\pm$0.35  & 2290$\pm$77  & 0.44$\pm$0.14  & 1.55$\pm$0.09  \\
2165-53917-0009 & 150$\pm$28   & 4359$\pm$524  & 0.478$\pm$0.024 
	& 1.36$\pm$0.05  & 1030$\pm$57  & 0.06$\pm$0.01  & 2.42$\pm$0.04  \\
2419-54139-0083 & 746$\pm$224   & 11193$\pm$3537  & 0.655$\pm$0.104 
	& 5.77$\pm$2.61  & 880$\pm$15  & 0.79$\pm$0.01  &  3.77$\pm$0.02 \\
2477-54058-0014 & 181$\pm$36   & 6080$\pm$946  & 0.522$\pm$0.037 
	& 1.24$\pm$0.04  & 1175$\pm$48  & 0.11$\pm$0.01  & 2.33$\pm$0.06  \\
2593-54175-0208 & 122$\pm$21   & 2484$\pm$297  & 0.416$\pm$0.018 
	& 1.29$\pm$0.07  & 747$\pm$52  & 0.50$\pm$0.01  & 2.19$\pm$0.02  \\
2756-54508-0150 & 168$\pm$20   & 40329$\pm$31137  & 0.252$\pm$0.009 
	& 2.55$\pm$0.04  & 678$\pm$72  & 0.16$\pm$0.05  & 2.18$\pm$0.04  \\
2760-54506-0022 & 159$^{**}$   & 21907$^{**}$  & 0.857$\pm$0.767 
	& 0.84$\pm$0.11  & 1010$\pm$53  & 0.02$\pm$0.02  & 1.57$\pm$2.16  \\
2768-54265-0402 & 201$\pm$29   & 6911$\pm$930  & 0.463$\pm$0.024
	& 1.59$\pm$0.04  & 772$\pm$35  & 0.14$\pm$0.01  & 2.23$\pm$0.01  \\
2783-54524-0517 & 114$\pm$14   & 21989$\pm$2727  & 0.591$\pm$0.034 
	& 1.46$\pm$0.02  & 800$\pm$39  & 0.00$^{**}$  & 2.51$\pm$0.01  \\
\hline
\end{tabular}\\
Notice: $^{**}$ means the model determined value quite smaller than the corresponding uncertainties. \\
The first four rows are the results for the four double-peaked BLAGN from the sample of \citet{st03}. 
The fifth row to the fifteenth row are for the double-peaked BLAGN collected from SDSS quasars. 
The other rows are for the double-peaked BLAGN collected from the sample of \citet{liu19}\\
The listed $\sin(i)$ in the fourth column are the same as the ones listed in the seventh column of Table~1.
\end{table*}

\begin{table}
\caption{Line fluxes of broad Balmer emission lines of the 38 double-peaked BLAGN}
\begin{tabular}{ccccc}
\hline\hline
	PMF  & $f_{\beta}$  & $f_\alpha$  &  rat & $E(B-V)$  \\
\hline
	0721-52228-0600  &       588   &       3246    &      5.52  & 0.500  \\
	0725-52258-0510   &      19064 &         96874 &         5.08 &  0.428 \\
	0746-52238-0409   &      3086    &      14595 &         4.72  & 0.365 \\
	6139-56192-0440   &      1167    &      6299   &       5.39 &  0.480 \\
\hline\hline
	0749-52226-0411  &       1025   &       4908   &       4.78 &  0.375 \\
	1073-52649-0132   &      6527 &         15850  &        2.42 &  0.000 \\
	1393-52824-0216   &      811    &      6096  &         7.51 &  0.765 \\
	1618-53116-0082   &      1887    &      8411   &       4.45 &  0.312 \\
	1624-53386-0032 &        4234   &       23628  &        5.57 &  0.506 \\
	2091-53447-0584   &      1651 &         8837   &       5.34  & 0.470 \\
	2503-53856-0527  &       1453   &       7974 &         5.48  & 0.487 \\
	2586-54169-0451   &      3238    &      14724  &        4.54 &  0.330 \\
	2655-54184-0450 &        1723   &       8707  &        5.05 &  0.420 \\
	2784-54529-0027   &      2556 &         8205  &        3.21  & 0.027 \\
	2974-54592-0133   &      1498   &       6813 &         4.54  & 0.330 \\
\hline\hline
	0492-51955-0273   &      327    &      1813   &       5.53 &  0.498 \\
	0581-52356-0463 &        1065  &        3714  &        3.48 &  0.100 \\
	0667-52163-0506   &      1203 &         6212  &        5.16 &  0.440 \\
	0784-52327-0594   &      575    &      2363 &         4.11 &  0.240 \\
	0827-52312-0023   &      12902   &       40474  &        3.13 &  0.010 \\
	0966-52642-0499 &        2055  &        7941   &       3.86  & 0.185 \\
	0972-52435-0520 &        2067 &         12542  &        6.06 &  0.576 \\
	1061-52641-0116  &       4079  &        16065 &         3.93 &  0.205 \\
	1196-52733-0639   &      1842  &        8686  &        4.71 &  0.360 \\
	1229-52723-0299    &     730   &       3565   &       4.88 &  0.395 \\
	1360-53033-0206    &     1827  &        8003  &        4.37 &  0.301 \\
	1618-53116-0280    &     532   &       1984   &       3.72 &  0.156 \\
	2022-53827-0286    &     2264  &        11870 &         5.24  & 0.449 \\
	2108-53473-0120    &     9630  &        45910 &         4.76  & 0.374 \\
	2123-53793-0443    &     19853 &         78485 &         3.95  & 0.210 \\
	2165-53917-0009    &     741   &       4097   &       5.52  & 0.495 \\
	2419-54139-0083    &     2581 &         24967 &         9.67   &0.985 \\
	2477-54058-0014       &2276   &       9908  &        4.35  & 0.295\\
	2593-54175-0208    &  950     &     4607     &     4.84  &0.385 \\
	2756-54508-0150   &    417    &      2238     &     5.36  & 0.470 \\
	2760-54506-0022  &      496   &       2269    &      4.57  & 0.331 \\
	2768-54265-0402 &        791  &        4022   &       5.08  & 0.425 \\
	2783-54524-0517 &       3379   &       14498  &        4.29 &  0.275 \\
\hline
\end{tabular}\\
Notice: The first column shows the information of PLATE-MJD-FIBERID, the second and the third column show the 
measured line flux of broad H$\beta$ and broad H$\alpha$ in the units of $10^{-17}{\rm erg/s/cm^2}$, the fourth 
column shows the flux ratio of broad H$\alpha$ to broad H$\beta$, the last column shows the calculated $E(B-V)$.\\
The first four rows are the results for the four double-peaked BLAGN from the sample of \citet{st03}. 
The fifth row to the fifteenth row are for the double-peaked BLAGN collected from SDSS quasars. The other rows 
are for the double-peaked BLAGN collected from the sample of \citet{liu19}.
\end{table}

     Then, Figure~\ref{res} shows the dependence of $R_{NLRs,u}$ on the [O~{\sc iii}] line luminosity 
of the 38 low redshift double-peaked BLAGN. Here, based on the intrinsic flux ratio of 
narrow H$\alpha$ to narrow H$\beta$ as 3.1, reddening effects have been corrected to calculate the 
intrinsic [O~{\sc iii}] line luminosity for the objects with flux ratios of narrow H$\alpha$ to narrow 
H$\beta$ larger than 3.1, based on the well applied Galactic extinction curve in \citet{fk99}. The 
applied values of $E(B-V)$ are also listed in Table~1. More interestingly, the upper limits of NLRs 
sizes $R_{NLRs,u}$ of the 38 double-peaked BLAGN are not apparently against the expected 
results from the empirical relation of type-2 AGN in the literature. Based on the linear regression 
result on the correlation between NLRs sizes and [O~{\sc iii}] luminosities of the type-2 AGN, the 90\% 
and 99.9999\% confidence bands are shown for the linear regression result. All the type-2 AGN with 
measured NLRs sizes shown in \citet{liu13} are lying within the 99.9999\% confidence bands. Whereas, 
there are 35 double-peaked BLAGN lying within the 99.9999\% confidence bands, besides the 
three objects SDSS 0581-52356-0463, SDSS 1196-52733-0639 and SDSS 2756-54508-0150 above the upper 
boundary of the 99.9999\% confidence bands. The three objects are shown in Figure~\ref{res} as solid 
purple circles.

	However, considering the quite large uncertainties of NLRs sizes and the NLRs sizes as upper 
limits, it is hard to conclude that SDSS 0581-52356-0463, SDSS 1196-52733-0639 and SDSS 2756-54508-0150 
do not obey the dependence of NLRs sizes on [O~{\sc iii}] luminosity from type-1 AGN. In the near 
future, large sample of low redshift double-peaked BLAGN in SDSS could provide further clues on whether 
are there different dependence of NLRs size on [O~{\sc iii}] luminosity between type-1 AGN and 
type-2 AGN. In the current stage, the simple conclusion can be well reached that there are the similar 
dependence of NLRs sizes on [O~{\sc iii}] luminosity between the type-1 AGN and the type-2 AGN, 
indicating no apparent clues to against the expected results from the unified model for different 
kinds of AGN.

\begin{figure*}[htp]
\centering\includegraphics[width = 18cm,height=6cm]{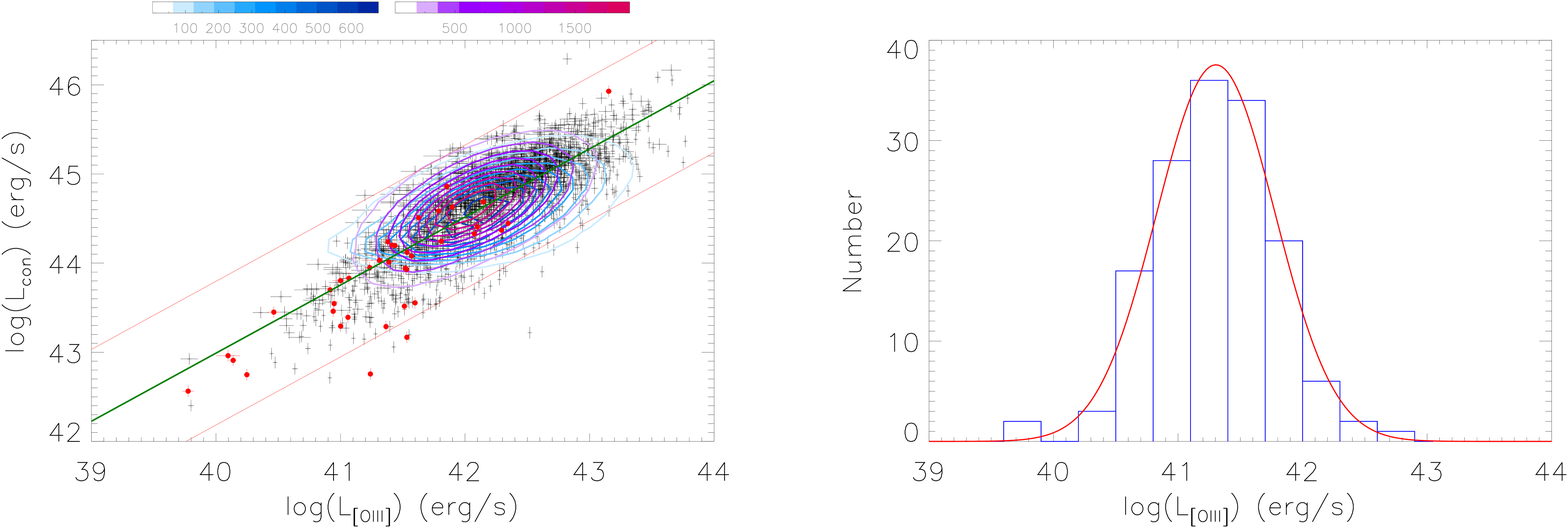}
\caption{Left panel shows the correlation between continuum luminosity and \o3~ luminosity. Solid dots 
plus error bars are the results for the type-1 AGN discussed in \citet{zh17b}. Contour in bluish colors 
represents the results for the quasars collected from \citet{sr10}, with the color bar in bluish colors 
shown in top corner of the left panel. Contour in reddish colors represents the results 
for the quasars collected from \citet{ra20}, with the color bar in bluish colors shown in top corner 
of the left panel. Solid red circles plus error bars are the results for the 38 low redshift double-peaked 
BLAGN in the final sample after intrinsic reddening corrections. Solid dark green line and solid red lines 
show the best-fitting results and corresponding 2.6RMS scatters to the correlation. Right panel shows the 
distributions of $\log(L_{[O~\textsc{iii}]})$ of the 149 Type-1 AGN with redshift smaller than 0.1\ in the 
sample of \citet{zh17b}.
}
\label{co3}
\end{figure*}

    Before further discussions on the results shown in Figure~\ref{res}, it is necessary to consider 
whether the SDSS fibers cover all the [O~{\sc iii}] emissions of the 38 low redshift double-peaked 
BLAGN? We answer the question as follows. As what we have known that there is one strong linear 
correlation between continuum luminosity and [O~{\sc iii}] luminosity of QSOs, see our previous results 
in \citet{zh17b} and results in \citet{sr10} and results in \citet{ra20}, if the SDSS fibers 
only cover part of [O~{\sc iii}] emissions, the observed [O~{\sc iii}] luminosity could be smaller 
than the expected value from the continuum luminosity, because the continuum emissions of type-1 
AGN are totally covered by the SDSS fibers. Therefore, we check the correlation between continuum 
luminosity and [O~{\sc iii}] luminosity of the 38 double-peaked BLAGN. Here, based on the intrinsic 
flux ratio of broad H$\alpha$ to broad H$\beta$ as 3.1, intrinsic reddening effects have been corrected 
to calculate the intrinsic continuum luminosity for the objects with flux ratios of broad H$\alpha$ to 
broad H$\beta$ larger than 3.1, based on the well applied Galactic extinction curve in \citet{fk99}. 
The line fluxes of the broad Balmer lines are calculated by sum of the fluxes of the broad Gaussian 
components shown as solid lines in Figure~\ref{line} and listed in Table~3. The applied $E(B-V)$ to 
correct continuum emissions are also listed in Table~3. The correlations are shown in left panel of 
Figure~\ref{co3} between the reddening corrected continuum luminosity and \o3~ line luminosity for 
the type-1 AGN discussed in \citet{zh17b} and for the quasars in \citet{sr10} and for the 
quasars in \citet{ra20}. Based on the shown results in left panel of Figure~\ref{co3}, two points can 
be found. On the one hand, the results in \citet{zh17b} are well consistent with those in \citet{sr10} 
and in \citet{ra20}, to support the robust correlation between continuum luminosity and 
\o3~ line luminosity in type-1 AGN. On the other hand, the observed [O~{\sc iii}] luminosities are 
not statistically smaller than the expected values through the continuum luminosities. Based on the 
results shown in the left panel of Figure~\ref{co3}, the mean values of continuum luminosity to \o3~ 
line luminosity $\log(L_{5100})-\log(L_{[O~\textsc{iii}]})$ are about 2.49$\pm$0.31, 
2.49$\pm$0.37, 2.57$\pm$0.41 and 2.51$\pm$0.36 for the type-1 AGN in \citet{zh17b}, for the quasars 
in \citet{sr10}, for the quasars in \citet{ra20}, and for the 38 double-peaked BLAGN, respectively, 
with the standard deviations as the uncertainties of the mean values. Student’s T-statistic technique 
is applied to confirm significance level higher than 75\% that the normal type-1 AGN and the 38 
double-peaked BLAGN have the same mean values of $\log(L_{5100})-\log(L_{[O~\textsc{iii}]})$. 
Therefore, the calculated $R_{NLRs,u}$ are truly the upper limits of sizes of NLRs of the 38 
double-peaked BLAGN, after accepted that the [O~{\sc iii}] emission regions are expected to be totally 
covered in SDSS fibers.

	Before given the final conclusions, it is necessary to check whether randomly collected 
values of $\sin(i)$ can lead to similar results as those in Figure~\ref{res} for the final sample 
of the 38 low redshift double-peaked BLAGN. A simple procedure is applied as follows. There are 500 
values of $\sin(i)$ randomly created with inclination angle $i$ from 20\degr~ to 70\degr, 500 values 
of redshift randomly created from 0 and 0.1. For the [O~{\sc iii}] line luminosity 
$\log(L_{[O~\textsc{iii}]})$, the values are created as follows. Based on the reported 
$\log(L_{[O~\textsc{iii}]})$ in \citet{zh17b} of the 149 low-redshift Type-1 AGN with $z<0.1$, 
distribution of $\log(L_{[O~\textsc{iii}]})$ of the 149 quasars shown in right panel of Figure~\ref{co3} 
has a Gaussian profile with mean value of 41.3 and standard deviation of 0.48. Although \citet{sr10} 
and \citet{ra20} have reported large samples of SDSS quasars, there are only 21 quasars with redshift smaller 
than 0.1\ in the sample of \citet{sr10}, and only 53 quasars with redshift smaller than 0.1\ in the sample 
of \citet{ra20}. Therefore, there are no further considerations of the low redshift quasars in \citet{sr10, ra20} in 
right panel of Figure~\ref{co3}. Based on the Gaussian distribution of $\log(L_{[O~\textsc{iii}]})$ for 
quasars with $z<0.1$, 500 values of $\log(L_{[O~\textsc{iii}]})$ are created. Then, based on the 500 created 
$\log(L_{[O~\textsc{iii}]})$ and the 500 randomly created redshift and $\sin(i)$, the simulating dependence 
of upper limits of NLRs sizes on [O~{\sc iii}] line luminosity is shown as blue open circle in Figure~\ref{res}. 
It is clear that there are more than one third of the 500 simulating data points lying out of the 99.9999\% 
confidence interval of the expected results from the empirical relation between NLRs sizes and \o3~ 
luminosity in the type-2 AGN. However, among the 38 low-redshift double-peaked BLAGN in our final sample, 
only three objects lying out of the 99.9999\% confidence interval. Therefore, the results of the 38 
double-peaked BLAGN in Figure~\ref{res} have real physical properties.

   Moreover, as we have discussed in Section 2, the long-term variabilities of the 38 
double-peaked BLAGN should be checked, in order to ignore the probable BBHs model applied to describe 
the double-peaked broad H$\alpha$. The long-term light curves have been collected from the CSS and 
shown in Figure~\ref{lmc} for the 37 double-peaked BLAGN, besides the SDSS 2123-53793-0443 without 
light curve provided by the CSS. The commonly accepted generalized Lomb-Scargle technique (GLS) 
\citep{ln76, sj82, zk09, zb16} is applied to check the probable QPOs signs. Here, as 
shown in Figure~\ref{lmc}, there are no signs for apparent QPOs in the 37 double-peaked 
BLAGN. So that, there are no further discussions on the results in Figure~\ref{lmc}, and the BBH model 
is not preferred in the collected double-peaked BLAGN. The results shown in Figure~\ref{res} are the 
basic and natural results on the upper limits of NLRs sizes after considerations of inclinations.

\begin{figure*}[htp]
\centering\includegraphics[width = 18cm,height=19cm]{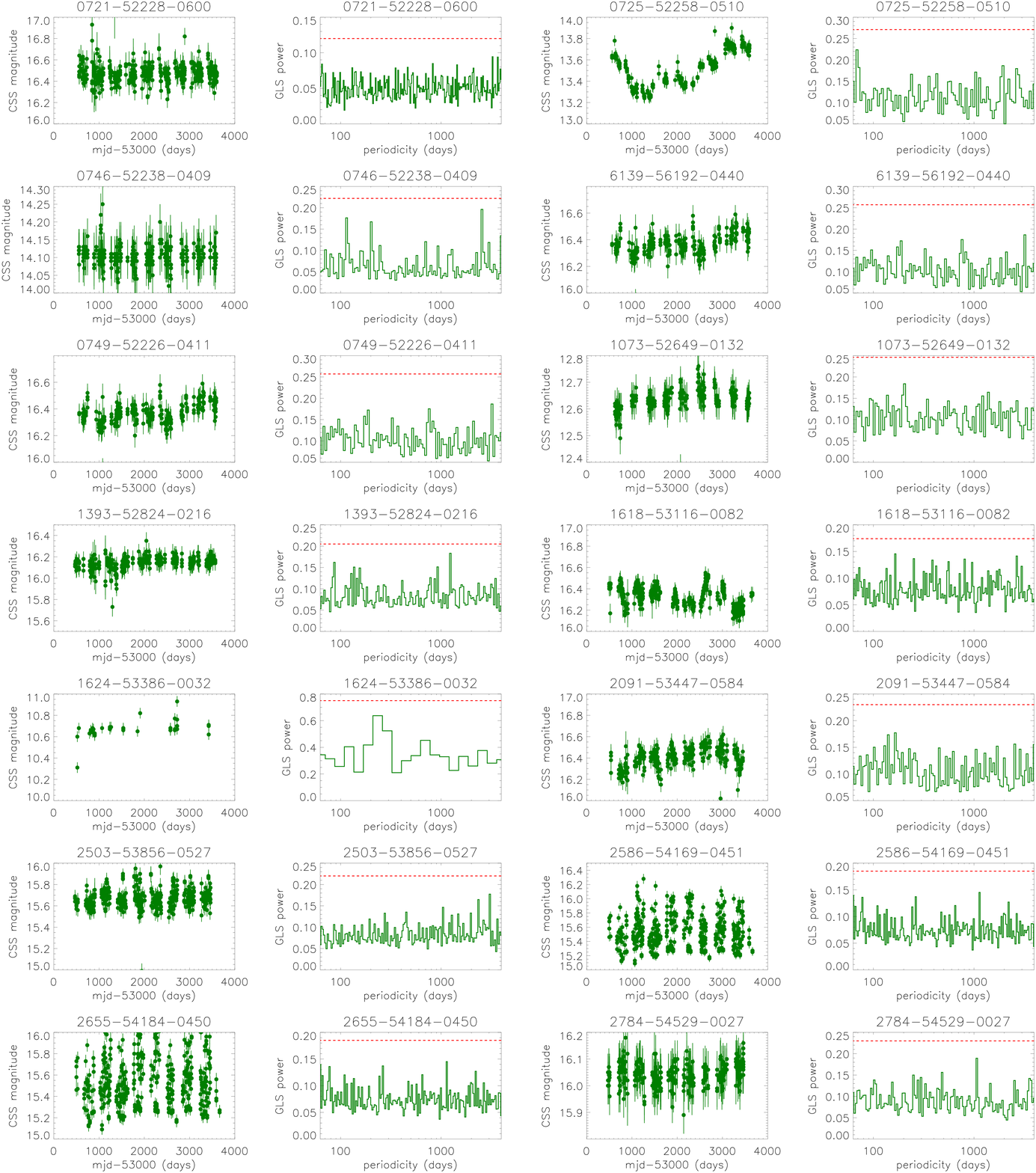}
\caption{Light curves of the 37 double-peaked BLAGN (panels in the first 
and the third column) and the corresponding GLS properties applied to detect QPOs signs ( 
panels in the second and the fourth column). In each panel for the GLS power properties, horizontal 
dashed red line shows the 99.99\% confidence level (0.0001 as the false-alarm probability) for the 
probable periodicity.}
\label{lmc}
\end{figure*}

\setcounter{figure}{8}

\begin{figure*}[htp]
\centering\includegraphics[width = 18cm,height=20cm]{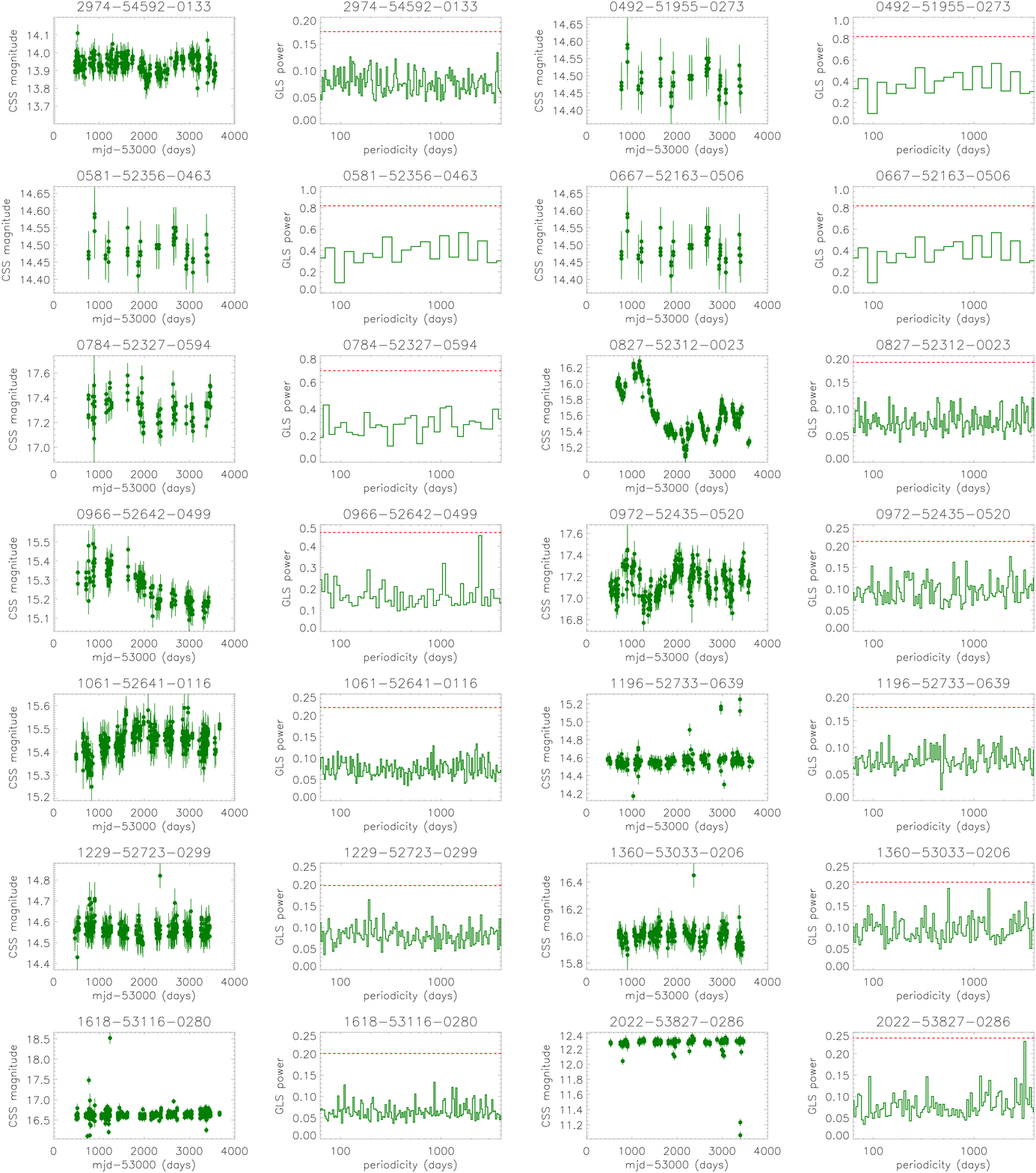}
\caption{--continued.}
\end{figure*}

\setcounter{figure}{8}

\begin{figure*}[htp]
\centering\includegraphics[width = 18cm,height=15cm]{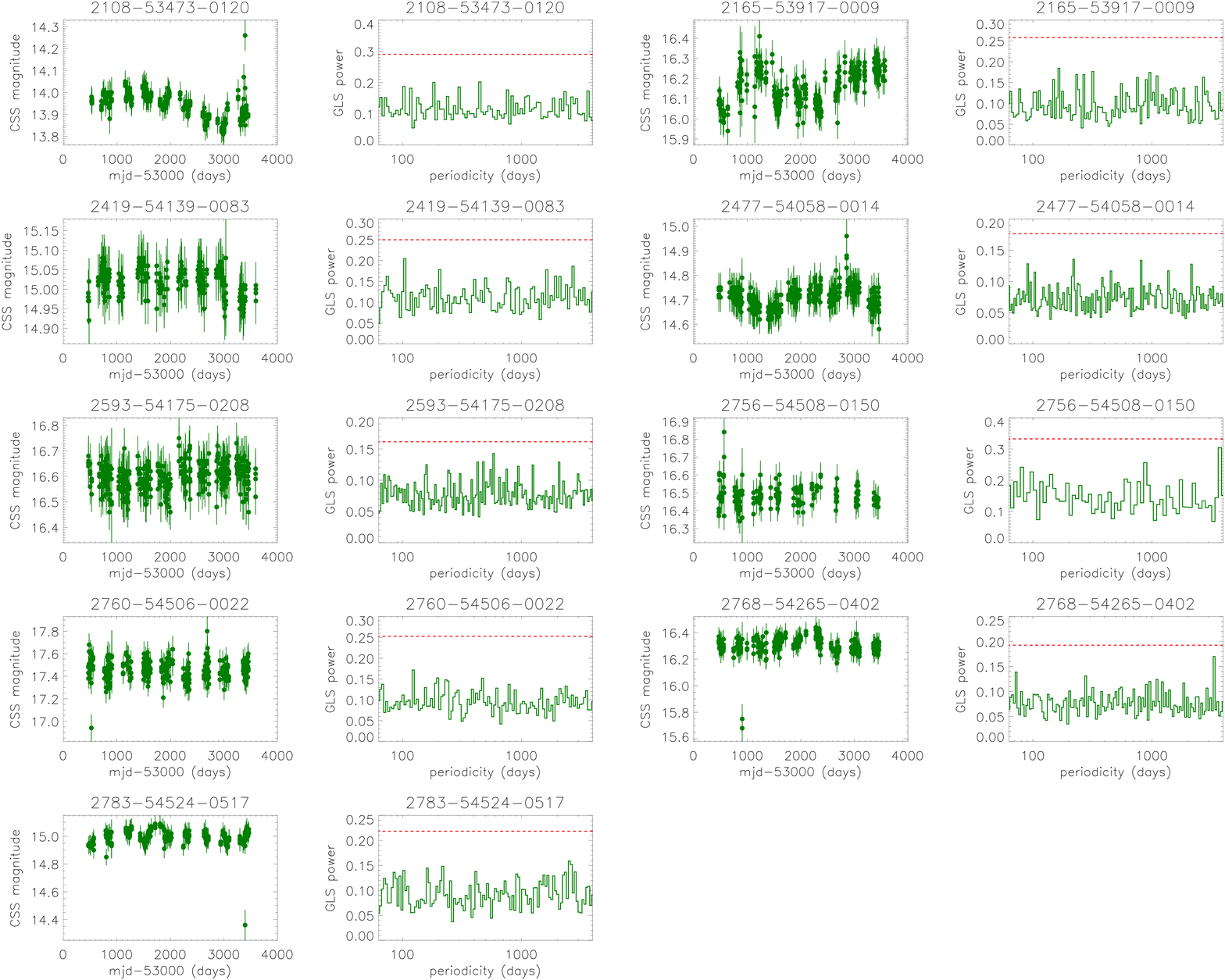}
\caption{--continued.}
\end{figure*}

\section{Summaries and Conclusions}

	Finally, we give our main summaries and conclusions as follows.
\begin{itemize}
\item  Double-peaked broad H$\alpha$ of 38 low redshift ($z<0.1$) double-peaked BLAGN can 
	be well described by the elliptical accretion disk model, accepted the accretion disk origin of 
	double-peaked broad emission lines leading to the well determined inclinations of central 
	line emission regions.
\item  Considering the fixed SDSS fibers and the determined inclinations, upper limits of NLRs sizes 
	in the 38 double-peaked BLAGN can be well estimated.
\item The strong linear correlation between continuum luminosity and [O~{\sc iii}] luminosity can be 
	well applied to confirm that the [O~{\sc iii}] emissions of the 38 low redshift 
	double-peaked BLAGN are totally covered in the SDSS fibers, indicating the estimated upper 
	limits of NLRs sizes are reliable.
\item There are no QPOs signs in the long-term light curves of the collected low redshift double-peaked 
	BLAGN, indicating the proposed BBH model is not preferred to explain the double-peaked broad 
	H$\alpha$ of the low redshift double-peaked BLAGN.
\item The 38 double-peaked BLAGN have their upper limits of NLRs sizes not statistically 
	against the expected results through the R-L relation for NLRs in type-2 AGN, indicating 
	that the current results can not provide clues to challenge the Unified model through the space 
	properties of NLRs.
\end{itemize}

\section*{Acknowledgements}
Zhang gratefully acknowledge the anonymous referee for giving us constructive comments 
and suggestions to greatly improve our paper. Zhang gratefully acknowledges the kind support of Starting 
Research Fund of Nanjing Normal University and from the financial support of NSFC-12173020. The manuscript 
has made use of the data from the SDSS (https://www.sdss.org/) funded by the Alfred P. Sloan 
Foundation, the Participating Institutions, the National Science Foundation and the U.S. 
Department of Energy Office of Science. The manuscript has made use of the long-term 
variability data from the CSS (http://nesssi.cacr.caltech.edu/DataRelease/).

\section*{Data Availability}
The data and program underlying this article will be shared on reasonable request to the corresponding 
author (\href{mailto:xgzhang@njnu.edu.cn}{xgzhang@njnu.edu.cn}).

\end{document}